\documentclass[a4paper,11pt]{article}
\pdfoutput=1 
\usepackage{jcappub}
\usepackage{amsmath}
\usepackage{graphicx}
\usepackage{latexsym}
\usepackage{xspace}
\usepackage{color}
\usepackage{hyperref} 
\usepackage{bm}
\usepackage{relsize}
\usepackage{tabularx}
\usepackage{multirow}
\usepackage{amssymb}
\usepackage[table]{xcolor}
\usepackage{braket}
\usepackage{comment}
\usepackage{slashed}

\makeatletter
\gdef\@fpheader{}
\g@addto@macro\bfseries{\boldmath}
\makeatother

\newcommand{\ie}{\textsl{i.e.~}}

\newcommand{\eg}{\textsl{e.g.~}}

\newcommand{\mean}[1]{\left\langle #1 \right\rangle}

\DeclareMathOperator{\erf}{erf}

\newcommand{\dd}{\mathrm{d}}
\newcommand{\ee}{e}

\newcommand{\sss}[1]{{\scriptscriptstyle{#1}}}

\newcommand{\uPl}{\mathrm{Pl}}
\newcommand{\uin}{\mathrm{in}}

\newcommand{\uc}{\mathrm{c}}

\newcommand{\usssPl}{\sss{\uPl}}

\newcommand{\Rea}{\Re \mathrm{e}\,}

\newcommand{\GeV}{\mathrm{GeV}}

\newcommand{\Mp}{M_\usssPl}

\newcommand{\efolds}{$e$-folds}
\newcommand{\efold}{$e$-fold}

\newcommand{\beq}{\begin{equation}}
\newcommand{\eeq}{\end{equation}}
\newcommand{\bea}{\begin{equation}\begin{aligned}}
\newcommand{\eea}{\end{aligned}\end{equation}}

\newlength{\wsingfig}
\newlength{\wdblefig}
\newlength{\wquadfig}
\newlength{\wtriplefig}
\newcommand{\Eq}[1]{eq.~(\ref{#1})}
\newcommand{\Eqs}[1]{eqs.~(\ref{#1})}
\newcommand{\Fig}[1]{fig.~{\ref{#1}}}
\newcommand{\Figs}[1]{figs.~{\ref{#1}}}
\newcommand{\Refa}[1]{ref.~{\cite{#1}}}
\newcommand{\Refs}[1]{refs.~{\cite{#1}}}
\newcommand{\Sec}[1]{sec.~\ref{#1}}

\newcommand{\App}[1]{appendix~\ref{#1}}

\newcommand{\f}{\phi}

\newcommand{\fo}{\f_{\mathrm{0}}}
\newcommand{\M}{\Mp}
\newcommand{\h}{\mathcal{H}}
\newcommand{\z}{z}
\newcommand{\x}{x}
\newcommand{\y}{y}

\newcommand{\ze}{z_{\mathrm{end}}}

\newcommand{\ye}{y_{\mathrm{end}}}
\newcommand{\pie}{\pi_{\mathrm{end}}}

\newcommand{\xin}{x_{\mathrm{in}}}
\newcommand{\yin}{y_{\mathrm{in}}}
\newcommand{\zin}{z_{\mathrm{in}}}
\newcommand{\PDF}{P}
\newcommand{\G}{P^{\mathrm{NB}}}
\newcommand{\PFPT}{P_{\mathrm{FPT}}}
\newcommand{\PFPTpole}{P_{\mathrm{FPT}}^{\chi_{\widetilde{N}}}}
\newcommand{\GFPT}{P_{\mathrm{FPT}}^{\mathrm{NB}}}
\newcommand{\Pleft}{p_{\text{re-entry}}}
\newcommand{\PBW}{P_{\mathrm{BW}}}
\newcommand{\GBW}{P_{\mathrm{BW}}^{\mathrm{NB}}}
\newcommand{\chafunc}{\chi_{\widetilde{N}}}

\DeclareMathOperator\erfc{erfc}

\subheader{}

\title{Uphill inflation}

\author[a]{Vadim Briaud,}
\emailAdd{vadim.briaud@phys.ens.fr}

\author[a]{Vincent Vennin}
\emailAdd{vincent.vennin@phys.ens.fr}

\affiliation[a]{Laboratoire de Physique de l'Ecole Normale Sup\'erieure, ENS, CNRS, Universit\'e PSL, Sorbonne Universit\'e, Universit\'e Paris Cit\'e, 75005 Paris, France}

\date{today}

\begin{document}
\sloppy

\abstract{
Primordial black holes (PBH) may form from large cosmological perturbations, produced during inflation when the inflaton's velocity is sufficiently slowed down. This usually requires very flat regions in the inflationary potential. In this paper we investigate another possibility, namely that the inflaton climbs up its potential. When it turns back, its velocity crosses zero, which triggers a short phase of ``uphill inflation'' during which cosmological perturbations grow at a very fast rate. This naturally occurs in double-well potentials if the width of the well is close to the Planck scale. We include the effect of quantum diffusion in this scenario, which plays a crucial role, by means of the stochastic-$\delta N$ formalism. We find that ultra-light black holes are produced with very high abundances, which do not depend on the energy scale at which uphill inflation occurs, and which suffer from substantially less fine tuning than in alternative PBH-production models. They are such that PBHs later drive a phase of PBH domination.
}

\arxivnumber{2301.09336}

\maketitle



\section{Introduction}
\label{sec:intro}

In the early universe, the gravitational collapse of large density fluctuations may give rise to primordial black holes~\cite{Hawking:1971ei, Carr:1974nx, Carr:1975qj} (PBHs). Among the different mechanisms~\cite{Escriva:2022duf} that have been proposed to account for the presence of those large perturbations is the parametric amplification of quantum vacuum fluctuations during cosmic inflation~\cite{Starobinsky:1979ty, Mukhanov:1981xt, Starobinsky:1982ee, Guth:1982ec, Bardeen:1983qw}. This same mechanism is responsible for the cosmological perturbations observed in the cosmic-microwave-background anisotropies~\cite{Planck:2018vyg, Planck:2018nkj}, and in the large-scale structures of the universe~\cite{SDSS:2005xqv, BOSS:2014hwf, Amendola:2016saw, DES:2022qpf}. Those measurements confirmed that cosmological perturbations are almost scale-invariant, Gaussian and adiabatic, which is consistent with single-field models of inflation~\cite{Martin:2013tda, Planck:2018jri}.

If inflation is driven by a single scalar field, perturbations get large when the field velocity is small, which in general requires to cross very flat regions of the potential function. Typical examples involve inflection-points models~\cite{Choudhury:2013woa, Kawasaki:2016pql, Garcia-Bellido:2017mdw, Ezquiaga:2017fvi, Germani:2017bcs,  Ezquiaga:2018gbw, Bhaumik:2019tvl, Figueroa:2020jkf}, false-vacuum configurations~\cite{Dvali:2003vv, Hamada:2014wna, Bezrukov:2014bra, Kitajima:2019ibn, Geller:2022nkr, Gu:2022pbo, Animali:2022otk}, step-like or bump-like features~\cite{Atal:2019cdz, Atal:2019erb, Inomata:2021tpx, Cai:2021zsp, Cai:2022erk, Pi:2022ysn} or non-canonical kinetic terms~\cite{Chen:2013aj, Chen:2013eea}. Another possibility is to consider the case where the inflaton climbs up its potential: at some point, the field stops climbing and starts rolling down, hence at the turn-around point its velocity necessarily vanishes. When this happens, inflation must take place (since the kinetic energy of the inflaton vanishes), and we refer to this regime as ``uphill inflation''. 

Interestingly, the production of PBHs in such a scenario was studied early on in \Refa{Yokoyama:1998pt}, in double-well potentials similar to the one depicted in \Fig{fig:DWIpot}. If the width of the double well is of the order of the Planck mass, the velocity inherited by the inflaton from the first phase of inflation taking place at large-field values is such that, after overshooting the potential minimum, the inflaton climbs up the local maximum and its velocity vanishes close to the top, before it rolls down again towards one of the two minima. This triggers a second phase of inflation during which perturbations grow at a high rate and may thus later collapse into PBHs.

The configuration leading to maximal PBH production is the one where the inflaton comes to a rest exactly at the potential maximum. Classically, it would inflate there forever. However, quantum fluctuations are expected to carry the system away from such an unstable equilibrium configuration. This indicates that the backreaction of quantum fluctuations onto the background dynamics plays an important role in this model, and the goal of this paper is to carefully study this effect. 
If large fluctuations are produced, such a backreaction effect is known to be crucial in shaping their statistical properties~\cite{Pattison:2017mbe, Ezquiaga:2019ftu, Panagopoulos:2019ail, Figueroa:2020jkf, Pattison:2021oen, Achucarro:2021pdh, Hooshangi:2021ubn, Ezquiaga:2022qpw, Cohen:2022clv} so this is also why it must be properly accounted for. We do so by using the stochastic-$\delta N$ formalism~\cite{Starobinsky:1982ee, Starobinsky:1986fx, Fujita:2013cna, Vennin:2015hra}, in which small-scales quantum fluctuations shift the background dynamics when they are stretched to large distances, and effectively act as a random noise. 

We find that the abundance of PBHs takes a universal value that is always large and that does not depend on the energy at which uphill inflation occurs. Since it takes place over the last few \efolds~of inflation, black holes are produced with very small masses that Hawking evaporate before big-bang nucleosynthesis, but they are so abundant that they drive a phase of PBH domination in the early Universe. The amount of fine tuning required for this scenario to take place is substantially smaller than in alternative models, which makes it attractive. 

In passing, when employing the stochastic-$\delta N$ formalism we propose two technical improvements, which are developed in the context of the present model but that are nonetheless of broader applicability. The first one concerns the implementation of the boundary conditions in the first-passage-time problem, where we find an approximation scheme that proves impressively efficient when the end-of-inflation surface is located in the drift-dominated region of phase space. The second one is a computation of the convolution integrals derived in \Refa{Tada:2021zzj} for the one-point statistics of the coarse-grained curvature perturbations, by means of a sampling algorithm that is not more expensive than a mere first-passage-time calculation. This should be useful for future implementations of the stochastic-$\delta N$ approach in other contexts.

The rest of this paper is organised as follows. In \Sec{sec:double-well inflation}, we present the scenario of uphill inflation at the classical level, both for the background and for linear cosmological perturbations. In \Sec{sec:stochastic formalism} we show how quantum diffusion can be included in the model, and we derive the relevant stochastic equations. In \Sec{sec:Fokker Planck} we solve the first-passage time problem associated to these stochastic equations, which leads us to the abundance of PBHs from uphill inflation in \Sec{sec:BW and PBH}. We present our conclusions in \Sec{sec:Conclusions}, and defer to two appendices some of the technical details that are not essential to the understanding of the main arguments developed in the main text. 
\section{Uphill inflation}
\label{sec:double-well inflation}
We consider the situation where the inflaton climbs up a hilltop potential. As the field slows down, a phase of inflation may take place before the inflaton falls off the local maximum. Such a scenario is referred to as ``uphill inflation''.
\subsection{Double-well realisation}
\label{sec:double:well}
\begin{figure}
\centering
\includegraphics[width=0.50\textwidth]{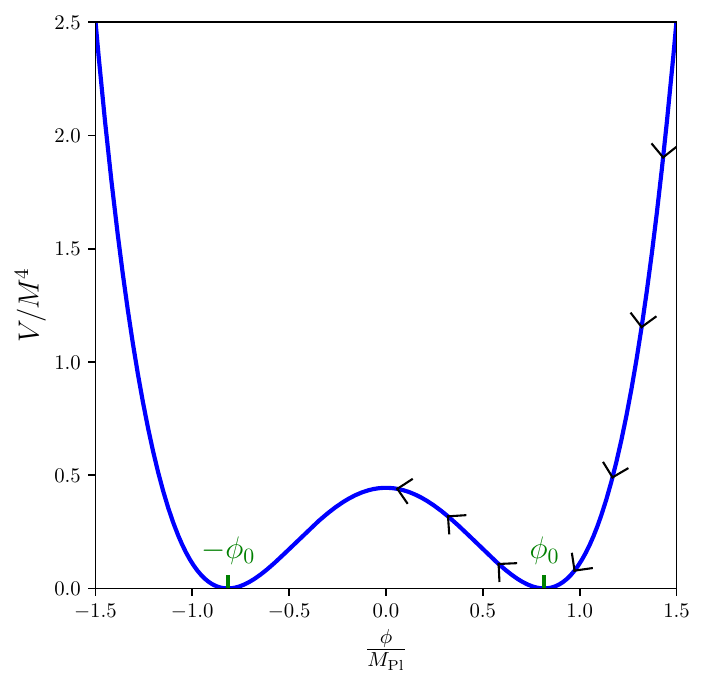}
\includegraphics[width=0.48\textwidth]{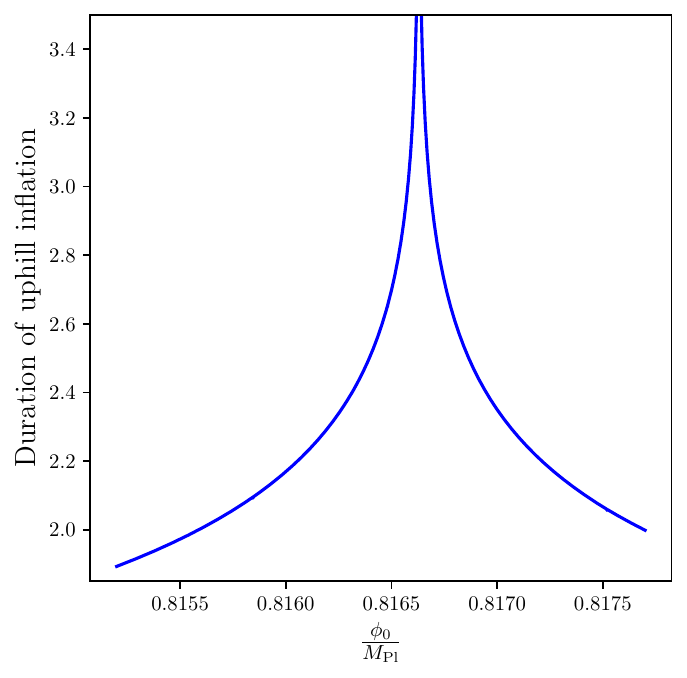}
			\caption{Left panel: double-well potential~\eqref{eq:pot:dwi} for $\fo=0.8166\Mp$. Right panel: Number of inflationary \efolds~realised between the first two crossings of a local minimum, as a function of $\fo$.} 
			\label{fig:DWIpot}
\end{figure} 
Although the results presented in this work do not depend on the detailed way uphill inflation is realised, for concreteness we first describe the case where it occurs in a double-well potential of the form
\beq
\label{eq:pot:dwi}
V(\f) = M^4\left[\left(\frac{\f}{\Mp}\right)^2-\left(\frac{\fo}{\Mp}\right)^2\right]^2\, ,
\eeq
which is displayed in the left panel of \Fig{fig:DWIpot}. In this model, inflation proceeds from large-field values, say at decreasing $\f$ in the direction indicated by the arrows in \Fig{fig:DWIpot}. When the inflaton approaches its local minimum, inflation stops by violation of the slow-roll conditions. Details about this first phase of slow-roll inflation can be found for instance in \Refa{Martin:2013tda} (under the name ``double-well inflation''). What matters is that since slow roll is a dynamical attractor, it quickly erases any dependence on the initial field value and its velocity. The inflaton then crosses the local minimum and starts to climb up the hill. 

Whether or not it overshoots the local maximum depends only on the value of $\phi_0$. The reason is the following. The classical dynamics of the inflaton is driven by the Klein-Gordon and Friedmann equations
\bea 
\ddot{\phi}+3H\dot{\phi}+\frac{\dd V}{\dd\f}=0
\quad\quad\text{and}\quad\quad
H^2=\frac{V+\frac{\dot{\phi}^2}{2}}{3\Mp^2}\, ,
\eea
where $H=\dot{a}/a$ is the Hubble parameter and a dot denotes derivation with respect to cosmic time. In terms of the number of \efolds~$N=\ln a$, they can be written  as~\cite{Chowdhury:2019otk} 
\beq
\label{eq:KG}
\frac{2}{6\Mp^2-\pi^2}\frac{\dd\pi}{\dd N}+\frac{\pi}{\Mp^2}+\frac{\dd\ln V}{\dd\phi}=0
\quad\quad\text{and}\quad\quad
H^2=\frac{V}{3\Mp^2-\frac{\pi^2}{2}}
\, ,
\eeq
where we have introduced $\pi\equiv\dd\f/\dd N$. 
The Klein-Gordon equation for $\f(N)$ does not involve $M^4$, and since initial conditions are washed out by the preceding dynamical attractor, the only relevant parameter of the (classical version of the) model is indeed $\fo$. Upon solving \Eq{eq:KG} numerically, we find that the inflaton falls on the other side of the hill if $\fo<\fo^\mathrm{cri}$, where
\beq
\label{eq:phi0:cri}
\fo^\mathrm{cri}\simeq 0.8166\Mp\, .
\eeq 
Otherwise, it turns over and remains on the same side of the hill. 

When the inflaton climbs up its potential, its velocity decays and the first Hubble-flow parameter 
\beq
\label{eq:eps1:def}
\epsilon_1\equiv -\frac{\dot{H}}{H^2}=\frac{\pi^2}{2\Mp^2}
\eeq
may become smaller than one (this is necessarily the case if $\fo>\fo^{\mathrm{cri}}$, since then the inflaton's velocity vanishes when it turns over, hence $\epsilon_1=0$ at that point). In that case, a second phase of inflation takes place, the duration of which is displayed in \Fig{fig:DWIpot}. One can check that when $\fo$ is fine-tuned to values close enough to $\fo^\mathrm{cri}$, the number of inflationary \efolds~in the uphill phase becomes large, and that it diverges at $\fo=\fo^\mathrm{cri}$. This is because, when $\fo=\fo^\mathrm{cri}$, the velocity inherited from the preceding phase of slow-roll inflation is such that it brings the inflaton to an exact rest at the location of the local maximum, where it inflates forever. 

As noted in \Refa{Yokoyama:1998pt}, this stationary configuration is nonetheless unstable, hence after some time it shall be broken by quantum fluctuations.\footnote{In \Refa{Yokoyama:1998pt}, a slightly different potential than \Eq{eq:pot:dwi} is considered, but it is still of the double-well type and the same considerations apply.} Since those fluctuations develop around a flat point in the potential, they acquire large cosmological amplitudes, which may result in the production of primordial black holes. To properly describe this mechanism, one has to account for the fact that fluctuations destabilise the background, and we will show below that this backreaction effect can be incorporated within the formalism of stochastic inflation. The goal of this work is to apply this formalism to uphill-inflation models, in order to investigate the production of large cosmological perturbations at the end of inflation in such scenarios.
\subsection{Classical phase space}
\label{sec:Classical:Phase:Space}
Let us start by studying the classical dynamics of uphill models. In practice, we focus on the region of the potential that is close to its local maximum, where $V(\f)$ can be approximated by
\beq
\label{eq:pot:quad:max}
V(\f) = V_0 - \frac{m^2}{2}\f^2\, .
\eeq
In the double-well model~\eqref{eq:pot:dwi}, $V_0=M^4(\fo/\Mp)^4$ and $m=2M^2\fo/\Mp^2$ but we now consider \Eq{eq:pot:quad:max} as a generic parametrisation of any quadratic local maximum. We simply assume that the initial velocity of the inflaton is close to the one that would make it freeze at the top of the hill, whether this velocity is set by a preceding slow-roll attractor phase as in the model detailed in \Sec{sec:double:well} or by any other mechanism.

Close to the local maximum, the inflaton's velocity becomes negligible, $\pi\ll \Mp$, and the potential is approximately constant, $V(\phi)\simeq V_0\simeq 3\Mp^2H_0^2$. In this limit, \Eq{eq:KG} becomes linear, and it can be solved according to
\beq
\label{eq:KGsol:phi}
\frac{\phi(N)}{\Mp} = x_\uin\ee^{\left(\nu_0-\frac{3}{2}\right)N}+y_\uin\ee^{\left(-\nu_0-\frac{3}{2}\right)N}
\eeq
where $x_\uin$ and $y_\uin$ are two integration constants and 
\beq
\label{eq:nu0:def}
\nu_0=\frac{3}{2}\sqrt{1+\frac{4m^2}{9H_0^2}}\, .
\eeq 
Since $\nu_0>3/2$, the first branch is unstable while the second branch decays away at late time. Therefore, the critical trajectory that  freezes on the local maximum at late time is such that $x_\uin=0$ and $y_\uin=\phi(N=0)/\Mp$, and in the following we consider trajectories close enough to this configuration, \ie such that $x_\uin\ll \phi(N=0)/\Mp$. The parameter $x_\uin$ thus quantifies the deviation from the critical trajectory. In the context of the double-well model~\eqref{eq:pot:dwi}, as will be made explicit in \Sec{sec:Conclusions}, this is related to the deviation of $\fo$ from $\fo^\mathrm{cri}$. 

Let us note that, in this regime, the three terms in the Klein-Gordon equation are comparable if the curvature of the potential at the end of the first phase of inflation is of order $m^2$. This is true in general, since for the first phase of inflation to end by violation of slow roll around a local minimum,  $m$ and $H$ need to be of the same order of magnitude. Assuming that $H$ does not decrease substantially between the two phases of inflation, $m$ is thus of order $H_0$, hence $\nu_0$ is of order one. In the double-well model described in \Sec{sec:double:well}, this is indeed the case since $m=2\sqrt{3}H_0\Mp/\fo$, so $\nu_0\simeq 4.5$. Then, $\dd V/\dd\phi = -m^2\phi$, $3H\dot{\phi}=-3H_0^2(3/2+\nu_0)\phi$ and $\ddot{\phi}=H_0^2(3/2+\nu_0)^2\phi$ are indeed all of the same order. This means that uphill inflation proceeds neither in the slow roll regime ($\ddot{\phi}\ll 3H\dot{\phi},\dd V/\dd\phi$), nor in the ultra slow-roll regime ($\dd V/\dd\phi \ll \ddot{\phi}, 3H\dot{\phi}$). It constitutes a ``new'' regime of inflation where the field acceleration, the Hubble friction and the potential gradient all play equally important roles.

\begin{figure}[t]
\begin{center}
\includegraphics[width=0.7\textwidth]{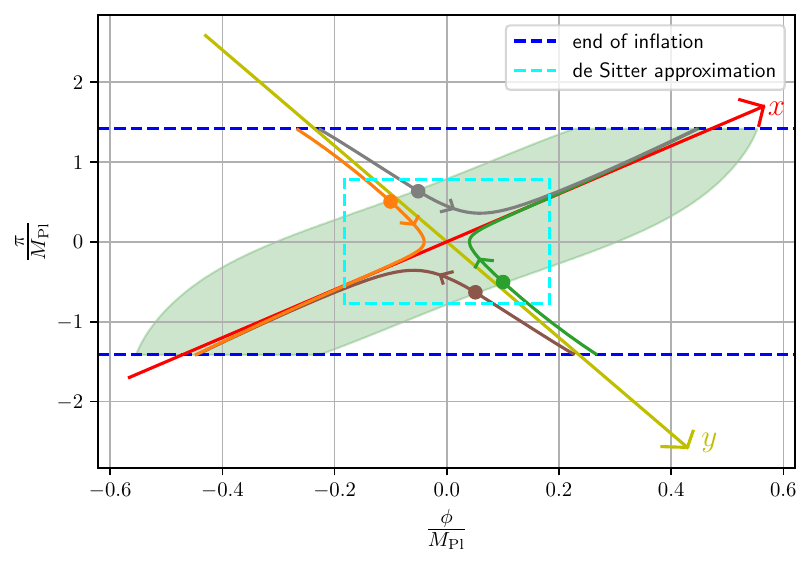}
\caption{Phase-space portrait of uphill inflation in the $\f,\dd\f/\dd N$ plane. The $ \x $ and $ \y $ directions introduced in \Eq{eq:x:y:def} are displayed in red and yellow respectively. Inflation takes place when $\vert \dd \f/\dd N\vert =\sqrt{2\epsilon_1}\Mp<\sqrt{2}\Mp$, \ie between the two dashed blue lines. The light-blue dashed rectangle corresponds to $m^2\f^2/2< V_0/10$, and $\left(\dd \f/\dd N\right)^2/2 < 3 \Mp^2/10$, which stands for the region where the first slow-roll parameter is small and the potential is almost constant, and is where the approximation~\eqref{eq:KGsol:phi} applies. This region is named ``de Sitter'' as $V$ and $H$ are almost constant (even though $ \epsilon_2 $ and $ \epsilon_3 $ are not small). The green region is where $ \frac{\delta\nu}{\nu_0} < 0.1 $, see also \Fig{fig:nu-tilde-phase-space-color-map}. A few trajectories are shown, on which coloured dots indicate the point where $\frac{\delta\nu}{\nu_0}$ drops below $0.1$ (\ie where the green region is entered). Trajectories are computed in the full potential~\eqref{eq:pot:dwi} where $\fo$ is given by its critical value~\eqref{eq:phi0:cri} and $M/\M=5\times 10^{-4}$.}  
\label{fig:Phase-space}
\end{center}
\end{figure}
The two branches appearing in \Eq{eq:KGsol:phi} also suggest that a more convenient parametrisation of phase may be provided by the variables $x$ and $y$ such that $\phi/\Mp=x+y$ and $\pi/\Mp=(\nu_0-3/2)x-(\nu_0+3/2)y$. These relations can be readily inverted and give
\bea 
\label{eq:x:y:def}
x&=\dfrac{1}{4\nu_0\Mp}\left[\left(3+2\nu_0\right)\phi+2\pi\right],\\
y&=-\dfrac{1}{4\nu_0\Mp}\left[\left(3-2\nu_0\right)\phi+2\pi\right].\\
\eea 
By inserting \Eq{eq:KGsol:phi} into \Eq{eq:x:y:def}, one obtains $x(N)=x_\uin \ee^{\left(\nu_0-3/2\right)N}$ and $y(N)=x_\uin \ee^{\left(-\nu_0-3/2\right)N}$, \ie $x$ and $y$ are the growing and decaying directions respectively. 

A phase-space portrait in the plane $(\f,\pi)$ is shown in \Fig{fig:Phase-space}, where the $x$ and $y$ directions are displayed in red and yellow respectively. Inflation takes place between the two dashed blue lines, which correspond to $\pi=\pm\sqrt{2}\Mp$ [\ie $\epsilon_1=1$, see \Eq{eq:eps1:def}]. The solution~\eqref{eq:KGsol:phi} applies when the field velocity can be neglected in the Friedmann equation $H^2=V/(3\Mp^2-\pi^2/2)$, and the potential is dominated by its constant term $V_0$. This region is delineated by the light-blue dashed rectangle, in which $\pi^2/2 < 3 \Mp^2/10$ and $m^2\f^2/2< V_0/10$. The $x$ and $y$ axes correspond to ``extremal'' trajectories: if initial conditions are set on the $x$ axis then the system remains on that axis and drifts away from the origin, while if initial conditions are set on the $y$ axis then the system is attracted towards the origin along that same axis. In this sense, the origin is an unstable saddle point.

\begin{figure}[t]
\begin{center}
\includegraphics[width=0.7\textwidth]{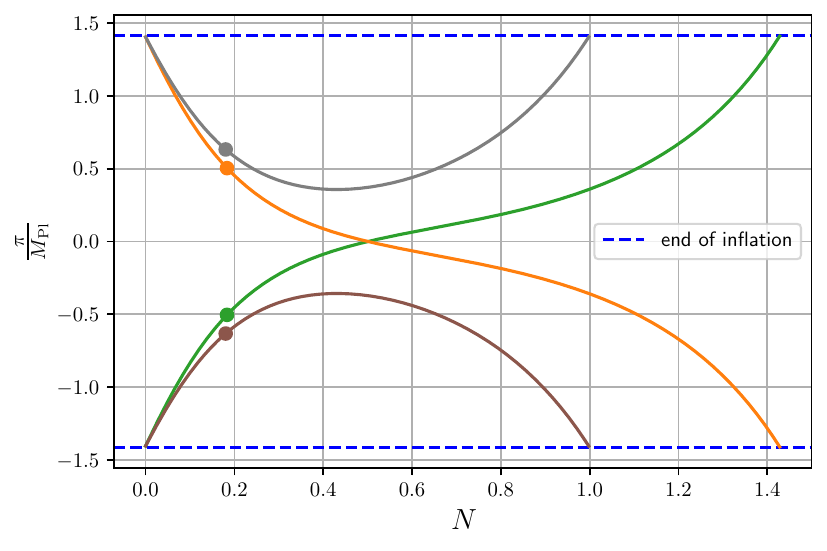}
\caption{Field velocity $\pi=\dd\f/\dd N$ as a function of the number of \efolds~$N$ for the four trajectories displayed in \Fig{fig:Phase-space}. As in \Fig{fig:Phase-space}, the blue dashed lines stand for $\epsilon_1=\pm 1$, and the coloured dots indicate the point where each trajectory enters the domain of phase space where $ \delta\nu/\nu_0 < 0.1 $. The origin $N=0$ for the number of \efolds~is set to the time when inflation resumes. }  
\label{fig:pi-trajectory}
\end{center}
\end{figure}
Four trajectories are represented in \Fig{fig:Phase-space}, one in each $(x,y)$ quarter plane. The green trajectory ($x>0,y>0$) climbs up the potential from positive field values, turns around before reaching the maximum and returns to the rightmost local minimum. Similarly, the orange trajectory ($x<0,y<0$) climbs up the potential from negative field values, turns around and returns to the leftmost local minimum. The brown (respectively grey) trajectory crosses over the local maximum leftwards (respectively rightwards). These four trajectories are also displayed in \Fig{fig:pi-trajectory}, where $\pi$ is shown as a function of $N$. 
\subsection{Linear cosmological perturbations}
\label{sec:CPT}
Let us now turn our attention to cosmological perturbations. This is required to compute the amplitude of the noise in the stochastic-inflation formalism. The scalar sector can be described by a single gauge-invariant combination, such as the Mukhanov-Sasaki variable $v(\vec{x},t)$~\cite{Mukhanov:1981xt,Kodama:1984ziu}. At linear order, its Fourier modes $v_k$ obey the  
Mukhanov-Sasaki equation
\beq
\label{eq:MS:eq}
v_k''+\left(k^2-\frac{Z''}{Z}\right) v_k=0\, ,
\eeq 
where a prime denotes derivation with respect to conformal time $\eta$, and $Z\equiv a\Mp\sqrt{2\epsilon_1}$. In terms of the Hubble-flow parameters, defined by $\epsilon_{n+1}= \dd\ln\epsilon_n/\dd N$, where $\epsilon_1$ was introduced in \Eq{eq:eps1:def}, one has
\beq
\frac{Z''}{Z} = \left(a H\right)^2\left(2-\epsilon_1+\frac{3}{2}\epsilon_2-\frac{1}{2}\epsilon_1\epsilon_2 + \frac{1}{4}\epsilon_2^2+\frac{1}{2}\epsilon_2\epsilon_3\right).
\eeq
Close to the phase-space origin, $H\simeq H_0$ hence $aH\simeq -1/\eta$, and $Z''/Z=(\nu^2-1/4)/\eta^2$ with $ \nu = \frac{3}{2}\sqrt{1+\frac{4}{9}(-\epsilon_1+\frac{3}{2}\epsilon_2-\frac{1}{2}\epsilon_1\epsilon_2 + \frac{1}{4}\epsilon_2^2+\frac{1}{2}\epsilon_2\epsilon_3)} $. The Hubble-flow parameters can be expressed in terms of $\phi$ and $\pi$ (or equivalently in terms of $x$ and $y$), hence they are phase-space functions. Indeed, from differentiating \Eq{eq:eps1:def} with respect to time and using the Klein-Gordon equation~\eqref{eq:KG}, one finds
\bea
\epsilon_2 &= \left(1+\frac{\Mp^2}{\pi}\frac{\dd\ln V}{\dd\f}\right)\left(\frac{\pi^2}{\Mp^2}-6\right),\\
\epsilon_3 &= \Mp^2\frac{\dd^2\ln V}{\dd\f^2}- \frac{\Mp^4  \frac{\dd^2\ln V}{\dd\f^2}}{\Mp^2+\frac{\pi}{\frac{\dd\ln V}{\dd \f}}}
+3\frac{\Mp^2}{\pi}\frac{\dd\ln V}{\dd \f} + \frac{\pi}{2}\frac{\dd\ln V}{\dd\f} +\frac{\pi^2}{\Mp^2}\, .
\eea
It is interesting to notice that these functions are not well defined at the origin of phase space $(\f=0,\pi=0)$, since they involve the ratio $(\dd\ln V/\dd\f)/\pi$, the value of which depends on the direction along which the origin is approached. If the inflaton climbs up its potential function and turns around at a point where the potential is not extremal, then this ratio (hence $\epsilon_2$ and $\epsilon_3$) even diverges. However, remarkably, this singular contribution cancels out in the combination giving $\nu$, and one finds
\bea
\label{eq:nu:phi:dphi_dN}
\nu=\frac{3}{2} &\left\lbrace
1+\frac{2}{9\Mp^4}\left[ \pi ^4+2\Mp^2 \frac{\dd\ln V}{\dd\f} \pi ^3+\Mp^2  \pi ^2\left(\Mp^2\frac{\frac{\dd^2 V}{\dd\f^2}}{V}-7\right)
\right.\right. \\ & \left. \left.
-12\Mp^4  \pi \frac{\dd\ln V}{\dd\f}-6\Mp^6\frac{\frac{\dd^2 V}{\dd\f^2}}{V} \right]
\right\rbrace^{1/2} .
\eea
This expression is perfectly regular when $\pi=0$, and at the phase-space origin it reduces to $\nu_0$, which was introduced in \Eq{eq:nu0:def}. The Mukhanov-Sasaki equation is therefore well-behaved everywhere in phase space,\footnote{One might be concerned with the terms involving $\dd\ln V/\dd\f$ and $\dd^2\ln V/\dd\f^2$ in \Eq{eq:nu:phi:dphi_dN}, which are singular when the potential vanishes. However, when $V=0$, the Friedmann equation \eqref{eq:KG} leads to $\pi=\sqrt{6}\Mp$, and these terms cancel out in \Eq{eq:nu:phi:dphi_dN}.} even though the higher Hubble-flow parameters are ill-defined in the configuration where the field comes to a rest.

\begin{figure}[t]
\begin{center}
\includegraphics[width=0.7\textwidth]{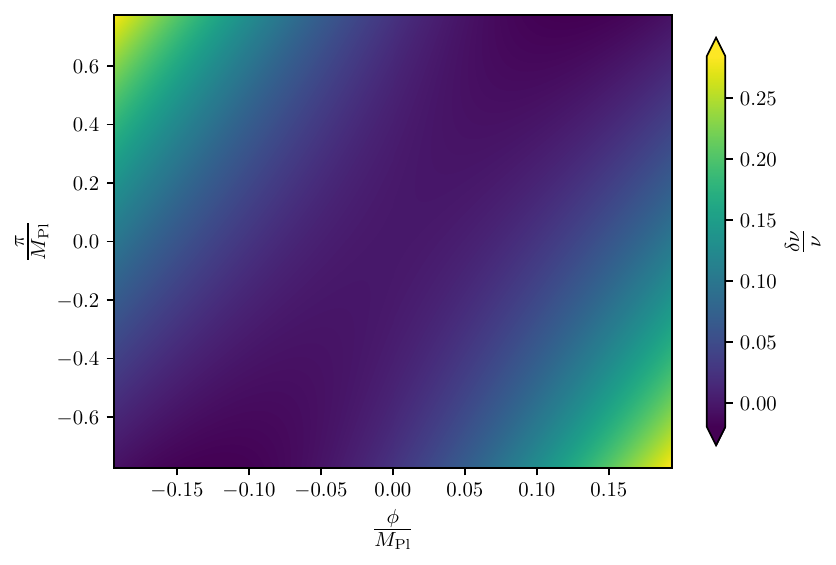}
\caption{Relative difference between the index $\nu$ and its value at the phase-space origin, $ \delta\nu/\nu_0=(\nu-\nu_0)/\nu_0 $, as a function of $\f$ and $\pi$. The boundaries of the axes match the dashed-blue rectangle in \Fig{fig:Phase-space}, inside of which $V$ and $H$ are almost constant.}  
\label{fig:nu-tilde-phase-space-color-map}
\end{center}
\end{figure}

The function $\nu(\f,\pi)$ given in \Eq{eq:nu:phi:dphi_dN} is shown in \Fig{fig:nu-tilde-phase-space-color-map}, where the relative difference $\delta\nu/\nu_0 = (\nu-\nu_0)/\nu_0$ is displayed. The frame of the figure corresponds to the dashed-blue rectangle in \Fig{fig:Phase-space}, within which $H\simeq H_0$ and $V\simeq V_0$ and our approximation applies. One can see that $\nu$ does not vary much within this domain, not more than $\sim 10\%$ except in the upper-left and bottom-right corners. In \Fig{fig:Phase-space}, the region where $\delta\nu/\nu_0<0.1$ is displayed in light green. One can check that as soon as a given trajectory enters that region (this moment is denoted by the coloured dots in \Figs{fig:Phase-space} and~\ref{fig:pi-trajectory}), it stays there until inflation ends. Moreover, once the system is attracted close to the $x$ axis (where we will see that most relevant stochastic effects take place), $\delta\nu/\nu_0$ is subject to tiny variations only, see \Fig{fig:nu-tilde-phase-space-color-map}. 

For these reasons, it is a good approximation to assume that $\nu\simeq \nu_0$ in the phase-space region of interest, and we have checked the validity of this approximation numerically by comparing the solution to \Eq{eq:MS:eq} with and without replacing $\nu=\nu_0$. 
One may be worried that, in practice, the system still inflates after exiting the domain where $H\simeq H_0, V\simeq V_0$ (the dashed blue rectangle in \Fig{fig:Phase-space}). However, this last stage of inflation is found to be very short-lived and subject to negligible stochastic effects, since it corresponds to the phase where the field is classically drifted away from the potential maximum. The loss of accuracy in the calculation of the noise amplitude in that phase has therefore negligible impact.

If $\nu$ is constant and equal to $\nu_0$, the Mukhanov-Sasaki equation~\eqref{eq:MS:eq} can be solved analytically in terms of the Hankel function of the first kind $H^{(1)}_{\nu_0}$, 
\beq
\label{eq:vk:Hankel}
v_k(\eta) = \frac{\sqrt{-\pi\eta}}{2} \ee^{i\frac{\pi}{2}\left(\nu_0+\frac{1}{2}\right)} H_{\nu_0}^{(1)}(-k\eta) \,,
\eeq 
where the integration constants have been set such as the Bunch-Davies vacuum $v_k(\eta)=\ee^{-ik\eta}/\sqrt{2k}$ is recovered in the sub-Hubble, asymptotic past. In the super-Hubble regime, $-k\eta\ll 1$, \Eq{eq:vk:Hankel} reduces to
\beq
\label{eq:v:superH}
v_k(\eta)\simeq -i \ee^{i\frac{\pi}{2}\left(\nu_0+\frac{1}{2}\right)}\frac{\Gamma(\nu_0)}{\sqrt{2\pi k}} \left(\frac{-k\eta}{2}\right)^{\frac{1}{2}-\nu_0}\, .
\eeq 
Let us note that, along the decaying branch $y$ [\ie when setting $x_\uin=0$ in \Eq{eq:KGsol:phi}], the curvature perturbations, $\zeta=v/Z$, quickly grows on super-Hubble scales. This is because $Z\propto \eta^{1/2+\nu_0}$ in that case, hence \Eq{eq:v:superH} implies that $\zeta\propto \eta^{-2\nu_0}$. With $\nu_0\simeq 4.5$ (corresponding to the double-well model described in \Sec{sec:double:well}), this leads to $\zeta\sim \eta^{-9}$, so the growth rate is indeed very large\footnote{This violates the bound derived in \Refa{Byrnes:2018txb}, which prevents super-$k^4$ growth of the power spectrum. This bound however only applies once transient regimes have died down. In our setup, the decaying branch $y$ eventually becomes smaller than the growing branch $x$, along which $Z\propto \eta^{1/2-\nu_0}$, hence $\zeta$ is frozen. This is expected as the $x$-branch is a dynamical attractor.} (recall that $\zeta$ is conserved in the presence of a phase-space attractor such as slow roll, and that it grows as $\zeta\sim \eta^{-3}$ in ultra slow-roll inflation). This suggests that large cosmological fluctuations develop in uphill inflation, that could be responsible for the formation of primordial black holes. These considerations however ignore the crucial role quantum diffusion plays in this model, which is the topic of the next section.

\section{Quantum diffusion}
\label{sec:stochastic formalism}

If the inflaton comes to a rest at a local maximum of its potential, it will soon be destabilised by quantum fluctuations, which implies that the classical description given in \Sec{sec:Classical:Phase:Space} breaks down. This can be modelled within the formalism of stochastic inflation, which we now briefly review. 
\subsection{Stochastic inflation}
Stochastic inflation~\cite{Starobinsky:1982ee, Starobinsky:1986fx} is an effective description for the long-wavelength part of quantum fields living on (and possibly sourcing) an inflating cosmological background. In practice, the fields are coarse-grained at the physical scale $R=(\sigma H)^{-1}$, where $\sigma\ll 1$ is the ratio between the Hubble radius and the coarse-graining radius, according to
\bea
\hat{\f}_{\mathrm{IR}}(\vec{x},N)&=\int\frac{\dd\vec{k}}{(2\pi)^{3/2}} W\left(\frac{k}{\sigma a H}\right)\left[ \ee^{-i \vec{k}\cdot\vec{x}}\f_{k}(N) \hat{a}_{\vec{k}}+\ee^{i \vec{k}\cdot\vec{x}}\f_{k}^*(N) \hat{a}^\dagger_{\vec{k}}\right]\, ,\\
\hat{\pi}_{\mathrm{IR}}(\vec{x},N)&=\int\frac{\dd\vec{k}}{(2\pi)^{3/2}} W\left(\frac{k}{\sigma a H}\right) \left[ \ee^{-i \vec{k}\cdot\vec{x}}\pi_{k}(N) \hat{a}_{\vec{k}}+\ee^{i \vec{k}\cdot\vec{x}}\pi_{k}^*(N) \hat{a}^\dagger_{\vec{k}}\right]\, .
\eea 
In this expression, $\hat{a}_{\vec{k}}$ and $\hat{a}_{\vec{k}}^\dagger$ are annihilation and creation operators, hats indicate that we are dealing with quantum operators, and $W$ is a window function that selects modes $k$ with wavelength larger than the coarse-graining scale, \ie $W\simeq 1$ for $k\ll \sigma a H$ and $0$ for $k\gg \sigma a H$. On super-Hubble scales, gradient terms can be neglected and the mode functions $\phi_k$ and $\pi_k$ follow the same equations of motion as the classical background. This is the so-called separate-universe approximation~\cite{Salopek:1990jq, Sasaki:1995aw, Wands:2000dp, Pattison:2019hef, Artigas:2021zdk}. The dynamics of $\hat{\f}_{\mathrm{IR}}$ and $\hat{\pi}_{\mathrm{IR}}$ is thus given by~\cite{Grain:2017dqa}
\bea 
\label{eq:Langevin:interm}
\frac{\dd\hat{\f}_{\mathrm{IR}}}{\dd N}&=\hat{\pi}_{\mathrm{IR}}+\hat{\xi}_\f(N)\, ,\\
\frac{\dd\hat{\pi}_{\mathrm{IR}}}{\dd N}&=-\left(3-\frac{\hat{\pi}_{\mathrm{IR}}^2}{2\Mp^2}\right)\hat{\pi}_{\mathrm{IR}}- \frac{\frac{\dd V}{\dd\f}\left(\hat{\f}_{\mathrm{IR}}\right)}{H^2\left(\hat{\f}_{\mathrm{IR}},\hat{\pi}_{\mathrm{IR}}\right)}+\hat{\xi}_\pi(N)\, ,
\eea 
where the source functions $\hat{\xi}_\f$ and $\hat{\xi}_\pi$ read
\bea 
\hat{\xi}_\f(N)&=-\int\frac{\dd\vec{k}}{(2\pi)^{3/2}} \left[ \ee^{-i \vec{k}\cdot\vec{x}}\f_{k}(N) \hat{a}_{\vec{k}}+\ee^{i \vec{k}\cdot\vec{x}}\f_{k}^*(N) \hat{a}^\dagger_{\vec{k}}\right]\frac{\dd }{\dd N}W\left(\frac{k}{\sigma a H }\right)\, ,\\
\hat{\xi}_\pi(N)&=-\int\frac{\dd\vec{k}}{(2\pi)^{3/2}} \left[ \ee^{-i \vec{k}\cdot\vec{x}}\pi_{k}(N) \hat{a}_{\vec{k}}+\ee^{i \vec{k}\cdot\vec{x}}\pi_{k}^*(N) \hat{a}^\dagger_{\vec{k}}\right]\frac{\dd }{\dd N}W\left(\frac{k}{\sigma a H }\right)\, ,
\eea 
and where the mode functions are to be evaluated in the uniform-$N$ gauge~\cite{Pattison:2019hef}.
The stochastic formalism then consists in treating \Eqs{eq:Langevin:interm} as stochastic, Langevin equations, where $\hat{\xi}_\f$ and $\hat{\xi}_\pi$ are replaced with stochastic noises, the statistics of which are drawn from their quantum expectation values. The noises $\hat{\xi}_\f$ and $\hat{\xi}_\pi$ involve small wavelength modes, which are described by standard cosmological perturbation theory. At linear order, they are placed in a Gaussian state, hence the noises are centred Gaussian noises. If $W$ is set to a Heaviside step function, \ie $W=1$ if $k<\sigma a H$ and $0$ otherwise, then their covariance is given by
\beq 
\label{eq:cov:interm}
\left\langle \hat{\xi}_f (N) \hat{\xi}_g^\dagger(N') \right\rangle = \frac{1}{6\pi^2} \frac{\dd \left(\sigma a H\right)^3}{\dd N} 
\Rea\left[f_{\sigma a H}(N) g^*_{\sigma a H}(N)\right]\delta(N-N')\, .
\eeq 
where $f,g=\f,\pi$. 

As mentioned above, $\f_k$ and $\pi_k$ correspond to the mode functions of the field fluctuations in the uniform-$N$ gauge [since perturbations to the lapse function have been discarded in \Eq{eq:Langevin:interm}]. In \Sec{sec:CPT}, we have explained how to compute the mode function of the Mukhanov-Sasaki variable, so these quantities now need to be related. The relevant relationship is obtained in \Refa{Pattison:2019hef}, and here we simply recall the result. 

The Mukhanov-Sasaki variable can be identified with the scalar field fluctuation in the spatially-flat gauge, $v_k/a=\f_k^\mathrm{flat}$. A gauge transformation is then needed to compute $\f_k$ in the uniform-$N$ gauge, which reads~\cite{Pattison:2019hef}
\beq
\label{eq:gauge:transform}
\phi_k = \f_k^\mathrm{flat}+ \f' \alpha_k\, ,
\eeq
where $\alpha$ is solution to the differential equation
\beq
\label{equa alpha}
3\h\alpha_k' + (3\h' -k^2)\alpha_k = S_k,
\eeq
where $\h=aH=a'/a$ is the conformal Hubble parameter, and the source $S_k$ is given by 
\beq
\label{source}
S_k = \frac{v_k\sqrt{2\epsilon_1}}{2 a\M}\mathrm{sign}(\dot\f)\left[\frac{\h\epsilon_2}{2}-\frac{(v_k/a)'}{v_k/a}\right]\, .
\eeq
Noting that \Eq{equa alpha} can be solved as
\beq
\label{formule alpha}
\alpha_k = \frac{1}{3\h}\int_{\eta_0}^{\eta}S_k(\eta')\exp\left[\frac{k^2}{3}\int_{\eta'}^{\eta}\frac{\mathrm{d}\eta''}{\h(\eta'')}\right]\dd\eta'\, ,
\eeq
the gauge transformation~\eqref{eq:gauge:transform} can be performed explicitly. In \Eq{formule alpha}, $\eta_0$ is an integration constant that defines the slicing relative to which the expansion is measured, which here we take in the asymptotic future ($\eta_0=0^-$).

In uphill inflation, \Eq{eq:v:superH} indicates that $v_k\propto \eta^{1/2-\nu_0}$ on super-Hubble scales, hence $(v_k/a)'/(v_k/a)=(2\nu_0-3)\h/2$.
Along the decaying branch $y$, one has $\epsilon_1\propto \eta^{3+2\nu_0}$, hence $\epsilon_2=-2\nu_0-3$. This leads to $S_k\propto \eta^2$ in \Eq{source}, hence $\alpha_k \propto \eta^4$ in \Eq{formule alpha}. The gauge correction in \Eq{eq:gauge:transform}, $\phi ' \alpha_k \propto \eta^{\nu_0+9/2}$, is therefore negligible compared to $\f_k^{\mathrm{flat}}=v_k/a\propto \eta^{3/2-v_0}$ on super-Hubble scales. Along the growing branch $x$,   one has $\epsilon_1\propto \eta^{3-2\nu_0}$, hence $\epsilon_2=2\nu_0+3$. This leads to an exact cancellation in the terms $\h \epsilon_2/2-(v_k/a)'/(v_k/a)$ appearing in the source function given in \Eq{source}. It implies that the gauge correction is even more suppressed in this case, which is expected since the growing branch acts as a dynamical attractor~\cite{Pattison:2019hef}.

We thus conclude that the gauge corrections are negligible on super-Hubble scales, where $\phi_k\simeq v_k/a \simeq 2^{\nu_0-1}H_0\Gamma(\nu_0)k^{-3/2}(-k\eta)^{3/2-\nu_0}/\sqrt{\pi }$ (up to an irrelevant global phase) and $\pi_k \simeq (\nu_0-\frac{3}{2})\f_k$. By substituting these expressions into \Eq{eq:cov:interm}, the covariance matrix of the noises is given by
\bea
\left\langle 
\begin{pmatrix}
\hat{\xi}_\f(N)\\
\hat{\xi}_\pi(N)
\end{pmatrix}
\begin{pmatrix}
\hat{\xi}_\f^\dagger(N') & \hat{\xi}_\pi^\dagger(N')
\end{pmatrix}\right\rangle=2\frac{\Mp^2}{\mu^2}
\begin{pmatrix} 1 &\ & \nu_0-\frac{3}{2}  \\ &\ & \\ \nu_0-\frac{3}{2} &\ & (\nu_0-\frac{3}{2})^2 \end{pmatrix}\delta\left(N-N'\right)\, ,
\eea 
where 
\beq 
\label{eq:mu:def}
\mu \equiv \frac{\sqrt{2}\pi^{3/2}}{\Gamma(\nu_0)}\frac{\Mp}{H_0}\left(\frac{\sigma}{2}\right)^{\nu_0-\frac{3}{2}}\, .
\eeq 
Since the determinant of the covariance matrix vanishes, there is a single independent noise, and one can write $\xi_\f=\sqrt{2}\Mp\xi/\mu$ and $\xi_\pi=(\nu_0-3/2)\sqrt{2}\Mp\xi/\mu$, where $\xi$ is a normalised white Gaussian noise, \ie such that $\langle \xi(N) \xi(N')\rangle = \delta(N-N')$. At the level of the approximation $\pi\ll \Mp$ performed here, the term proportional to $\hat{\pi}^2$ in \Eq{eq:Langevin:interm} must be neglected, and one obtains 
\bea
\label{Langevin}
    \frac{\dd\f}{\dd N} &= \pi + \sqrt{2}\frac{\Mp}{\mu} \xi (N) \, ,\\
    \frac{\dd\pi}{\dd N} &= -3\pi + \frac{m^2}{H_0^2}\f + \left(\nu_0-\frac{3}{2}\right)\sqrt{2}\frac{\Mp}{\mu} \xi (N)\, .
\eea
In this expression, the subscripts ``$\mathrm{IR}$'' have been dropped for notational convenience, and the hats have been removed too since one now deals with classical, random variables. These are the Langevin equations we aim at solving in the rest of this paper\footnote{Formally, when $m^2\to 0$ while keeping $H_0$ fixed, $\nu_0\to 3/2$, $\gamma\to H_0/(2\pi)$ and one recovers the ultra-slow roll Langevin equations~\cite{Pattison:2021oen}. This consistency check is nonetheless purely formal as ultra-slow roll is not a limiting case of our model (where $m$ and $H_0$ are of the same order).}.

These Langevin equations can also be rewritten in the $x$ and $y$ coordinates introduced in \Eq{eq:x:y:def}, and one obtains
\bea
\label{Langevin rotated}
    \frac{\dd\x}{\dd N} &= \left(\nu_0-\frac{3}{2}\right)\x + \frac{\sqrt{2}}{\mu}\xi (N) \, ,\\
    \frac{\dd\y}{\dd N} &= -\left(\nu_0+\frac{3}{2}\right)\y\, .
\eea
At the classical level (\ie without the stochastic noise $\xi$), these equations decouple, which is the reason why the variables $x$ and $y$ were introduced. We now find that these variables are also convenient at the stochastic level, since quantum diffusion only takes place along the $x$ direction. As a consequence, even if initial conditions are set along the $y$ direction (\ie such that the field classically freezes at the potential maximum), quantum diffusion explores the $x$ direction, and necessarily destabilises the system. We thus expect that, at the stochastic level, the number of inflationary \efolds~remains finite even when $x_\uin=0$, contrary to the classical setting (see the right panel of \Fig{fig:DWIpot}).
\subsection{The stochastic-$\delta N$ formalism}
\label{sec:StochaDeltaN}
Having a formalism at our disposal to describe the backreaction of quantum fluctuations onto the inflationary dynamics, the next step is to investigate how quantum backreaction affects observables, \ie the statistics of cosmological fluctuations. In standard cosmological perturbation theory, cosmological fluctuations are placed in a Gaussian state, the two-point functions of which are simply given by squaring the mode functions computed in \Sec{sec:CPT}. In stochastic inflation, the statistics of the curvature perturbation can be obtained using the so-called stochastic-$\delta N$ formalism~\cite{Enqvist:2008kt, Fujita:2013cna, Vennin:2015hra}. It relies on the fact that on super-Hubble scales, the local fluctuation in the amount of expansion $\mathcal{N}$ between an initial spatially-flat hypersurface and a final hypersurface of uniform energy density, is nothing but the curvature perturbation~\cite{Starobinsky:1982ee, Starobinsky:1986fxa, Sasaki:1995aw, Wands:2000dp, Lyth:2004gb}, $\zeta(\vec{x})=\mathcal{N}(\vec{x}) - \langle \mathcal{N} \rangle_{\vec{x}} =\delta\mathcal{N}(\vec{x})$. 

More precisely, when the curvature perturbation is coarse-grained at a scale $R$, the one-point statistics of $\zeta_R$ is given by the distribution function~\cite{Tada:2021zzj}
\beq
\label{eq:zeta:one:point:pdf}
P\left(\zeta_R\right) = \int \dd x_*\dd y_*\PBW\left[x_*,y_*\lvert N_{\mathrm{BW}}(R)\right]{\PFPT}_{x_\uin,y_\uin \rightarrow x_*,y_*}\left[\zeta_R-\langle\mathcal{{N}}\rangle\left(x_*,y_*\right)+\langle\mathcal{{N}}\rangle\left(x_\uin,y_\uin\right)\right] .
\eeq
In this expression, $N_{\mathrm{BW}}(R)$ corresponds to the number of \efolds~elapsed between the time when the scale $R$ exits the coarse-graining scale and the end of inflation (\efolds~need to be recorded backwards -- hence the subscript ``$\mathrm{BW}$'' -- from the end-of-inflation surface in order to correctly map $R$ to scales as measured by a local observer~\cite{Tada:2016pmk}). If $H$ is almost constant (which we assume here), it is given by $N_{\mathrm{BW}}(R)\simeq \ln(\sigma H R) $. The value of $x$ and $y$, $N_{\mathrm{BW}}(R)$ \efolds~before the end of inflation, are denoted $x_*$ and $y_*$ and they follow the ``backward'' distribution function~\cite{Ando:2020fjm}
\bea
\label{eq:PBW:def}
\PBW\left[x_*,y_*\lvert N_{\mathrm{BW}}(R)\right]= \PFPT\left[N_{\mathrm{BW}}(R)\vert x_*,y_*\right]\dfrac{\int_0^\infty \dd N' P\left(x_*,y_* \vert N',\xin,\yin\right)}{\int_{N_{\mathrm{BW}}(R)}^\infty \dd\mathcal{N} \PFPT\left(\mathcal{N}\vert x_\uin,y_\uin\right)}\, .
\eea 
This is the probability that, $N_{\mathrm{BW}}(R)$ \efolds~before the end of inflation, the value of $ \x $ and $ y $ is $x_*$ and $y_*$. Here, $\PFPT(N\vert x,y)$ is the probability that, starting from $x$ and $y$, inflation proceeds for $N$ \efolds~before ending. This is the so-called ``first-passage time'' (FPT) probability. Conversely, $P(x,y\vert N,\xin,\yin)$ is the probability that, starting from $x_\uin$ and $y_\uin$, after $N$ \efolds, the system is at field values $x$ and $y$. Finally, in \Eq{eq:zeta:one:point:pdf}, $\langle \mathcal{N}\rangle(x,y)$ denotes the mean first-passage time starting from $x$ and $y$, and ${\PFPT}_{x_\uin,y_\uin \rightarrow x_*,y_*}$ is the distribution for the time of first passage through $(x_*,y_*)$ (assuming this location is indeed visited at least once), starting from $(x_\uin,y_\uin)$. 

Similar expressions can be obtained for the two-point statistics, \ie the power spectrum (see \Refa{Ando:2020fjm}), and for the one-point statistics of the density contrast and of the compaction function (see \Refa{Tada:2021zzj}). Although seemingly not straightforward, they require to determine only two functions, $P$ and $\PFPT$, and in the next section we show how this can be done. 

When reconstructed numerically from Langevin simulations, $P(\zeta_R)$ is even simpler to obtain, along the following steps:
\begin{enumerate}
    \item \label{item:1} Pre-compute $\langle \mathcal{N} \rangle (x_*,y_*)$ on a grid of points $(x_*,y_*)$ to interpolate that function.
    \item \label{item:2} From the initial condition $(\xin,\yin)$, simulate one realisation of the Langevin equations and record the number of inflationary \efolds~$\mathcal{N}$ as well as the field values $(x_*,y_*)$ at a time $N_{\mathrm{BW}}$ before inflation ends.
    \item \label{item:3}Evaluate $\zeta_R = \mathcal{N}-N_{\mathrm{BW}}+\langle \mathcal{N}\rangle(x_*,y_*)-\langle \mathcal{N}\rangle(\xin,\yin)$ and store that value.
    \item Repeat steps \ref{item:2}-\ref{item:3} a large number of times.
    \item reconstruct $P(\zeta_R)$ from the obtained list of values for $\zeta_R$.
\end{enumerate}
This procedure allows one to sample $(x_*,y_*)$ according to the backward probability, hence the first-passage-time distribution is indeed weighted by that probability as required by \Eq{eq:zeta:one:point:pdf}. Note that, strictly speaking, the above steps sample ${\PFPT}_{x_\uin,y_\uin \rightarrow x_*,y_*}$ under the condition that between $(x_*,y_*)$ and the end-of-inflation surface, $N_{\mathrm{BW}}$ \efolds~are realised. However, the stochastic process being Markovian, the number of \efolds~realised before and after $(x_*,y_*)$ are independent, hence this additional condition is inoffensive. Apart from the pre-computation in step \ref{item:1}, this procedure is not computationally more expensive than the reconstruction of a mere first-passage time distribution.

\section{Field and first-passage-time probabilities}
\label{sec:Fokker Planck}

In this section, we derive the distribution functions $P$ and $\PFPT$, which respectively denote the probability associated to the field values after a given time, and the probability associated to the duration of inflation starting from given field values. It is first convenient to recast the Langevin equations~\eqref{Langevin rotated} in terms of the rescaled variables $ \z = \mu x \sqrt{(\nu_0-3/2)/2} $ and $ \widetilde{N} = (\nu_0-3/2 )N $,
\begin{align}
\label{Langevin rotated new variables:z}
    \frac{\dd\z}{\dd \widetilde{N}} &= \z + \xi (\widetilde{N})\, ,\\
    \frac{\dd\y}{\dd \widetilde{N}} &= -\frac{\nu_0+\frac{3}{2}}{\nu_0-\frac{3}{2}}\,\y\, .
\label{Langevin rotated new variables:y}
\end{align}
The Fokker-Planck equation driving $P(z,y\vert\widetilde{N},\zin,\yin)$ can then be derived~\cite{risken1989fpe},
\beq
\label{Fokker-Planck PDF}
\frac{\partial \PDF}{\partial \widetilde{N}} = \mathcal{L}_{\mathrm{FP}}\PDF,
\eeq
where $\mathcal{L}_{\mathrm{FP}}$ is the so-called Fokker-Planck operator and in the case of \Eqs{Langevin rotated new variables:z}-\eqref{Langevin rotated new variables:y} it is given by
\bea
\label{eq:LFP:def}
\mathcal{L}_{\mathrm{FP}} &= \frac{1}{2}\frac{\partial^2}{\partial \z^2} + \frac{\nu_0+\frac{3}{2}}{\nu_0-\frac{3}{2}}\, \frac{\partial}{\partial \y}\y - \frac{\partial}{\partial \z}\z \, .
\eea
Since the noise vanishes in the $ \y $-direction, $y$ follows a purely classical trajectory and $\mathcal{L}_{\mathrm{FP}}$ does not contain second derivatives with respect to $y$.
The Fokker-Planck equation~\eqref{Fokker-Planck PDF} needs to be solved with the initial condition $P(z,y\vert 0,\zin,\yin)= \delta(\z-\zin)\delta(\y-\yin) $, enforcing the fact that at initial time, the system starts at $\zin$ and $\yin$. Moreover, an absorbing boundary needs to be placed where inflation ends, \ie at $\pi=\pm \sqrt{2}\Mp$. In terms of the rescaled coordinates $z$ and $y$, this condition reads $2\sqrt{2\nu_0-3}z/\mu=(2\nu_0+3)y\pm2\sqrt{2}$, where we thus impose that $P$ vanishes at all times. 

Similarly, $\PFPT(\widetilde{N}\vert z,y)$ obeys the adjoint Fokker-Planck equation~\cite{Pattison:2017mbe}
\beq
\label{eq:adjoint:FP}
\frac{\partial \PFPT}{\partial \widetilde{N}} = \mathcal{L}_{\mathrm{FP}}^\dagger \PFPT\, ,
\eeq
where $ \mathcal{L}_{\mathrm{FP}}^\dagger$ is the adjoint of the operator introduced in \Eq{eq:LFP:def} and reads
\bea
\label{eq:adjoint:FP:operator}
\mathcal{L}_{\mathrm{FP}}^{\dag} &= \frac{1}{2}\frac{\partial^2}{\partial \z^2} -  \frac{\nu_0+\frac{3}{2}}{\nu_0-\frac{3}{2}}\, \y\frac{\partial}{\partial \y} + \z\frac{\partial}{\partial \z}\, .
\eea
When initial conditions are located on the end-of-inflation surface, the first-passage time vanishes, hence \Eq{eq:adjoint:FP} needs to be solved with the boundary condition $\PFPT(N\vert z,y)=\delta(N)$ when $2\sqrt{2\nu_0-3}z/\mu=(2\nu_0+3)y\pm2\sqrt{2}$.

\subsection{Field probability}
\label{sec:field:proba}

Although the two Langevin equations~\eqref{Langevin rotated new variables:z}-\eqref{Langevin rotated new variables:y} are uncoupled, the variables $z$ and $y$ are intertwined through the boundary condition defined on the end-of-inflation surface. This makes analytical treatments of the stochastic problem difficult to carry out. However, in practice, the role played by the coordinate $y$ can be neglected for the following reason.

\begin{figure}[t]
\begin{center}
\includegraphics[width=0.49\textwidth]{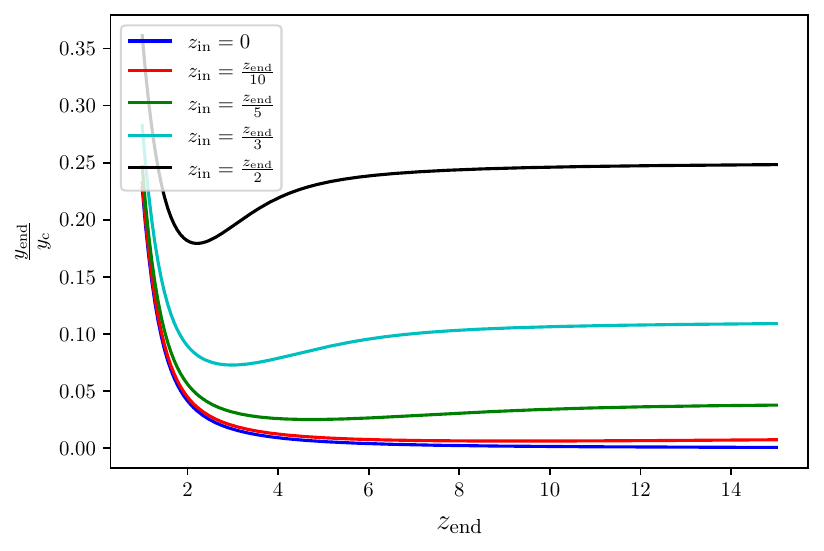}
\includegraphics[width=0.49\textwidth]{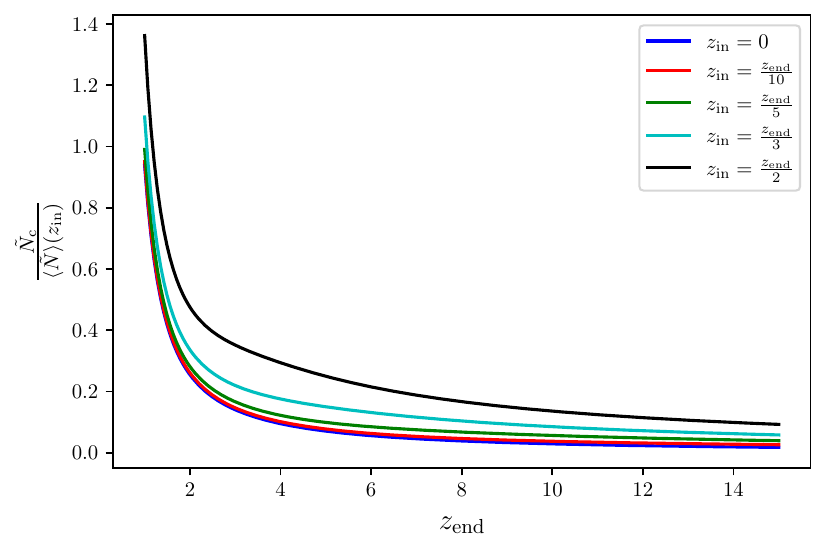}
\caption{Left panel: Relative error to the end-of-inflation condition carried by $y$, as a function of $ \ze $ and for a few values of $ \zin $, using the value $\nu_0\simeq 4.5$ of the double-well model~\eqref{eq:pot:dwi}.
Right panel: Comparison between the critical time $\widetilde{N}_\uc$ above which the return probability becomes small, and the mean duration of inflation $\langle\widetilde{N}\rangle(\zin)$, in the same situation as in the left panel.}
\label{fig:ratio time}
\end{center}
\end{figure}

The end-of-inflation condition can be rewritten as $z=\pm\ze(1\pm y /y_\uc)$, where $\ze = \mu/\sqrt{\nu_0-3/2}$ and $y_\uc=\sqrt{2}/(\nu_0+3/2)$. The correction $y$ carries in the boundary condition can thus be quantified by the ratio $\ye/y_\uc$. According to \Eq{Langevin rotated new variables:y}, $y$ decays exponentially in time, $\ye=\yin\ee^{-\widetilde{\mathcal{N}}(\nu_0+3/2)/(\nu_0-3/2)}$. In the double-well model introduced in \Sec{sec:double:well}, $\yin\simeq 0.24\Mp$, and the typical value for $\ye$ can be assessed by replacing $\widetilde{\mathcal{N}}$ by its expectation value in the above expression. From the adjoint Fokker-Planck equation~\eqref{eq:adjoint:FP}, one can show that the mean number of \efolds~starting from initial field values $\zin$ and $\yin$, $\langle \widetilde{\mathcal{N}}\rangle(\zin,\yin)=\int\PFPT(\widetilde{\mathcal{N}}\vert \zin,\yin) \widetilde{\mathcal{N}}\dd\widetilde{\mathcal{N}}$, obeys~\cite{Vennin:2015hra}
\beq
\label{mean number equation}
\mathcal{L}_{\mathrm{FP}}^{\dag}\langle \mathcal{\widetilde{N}}\rangle = -1\, ,
\eeq
with boundary condition $\langle\widetilde{\mathcal{N}}\rangle=0$ when starting on the end-of-inflation surface. Since our goal is to show that $y$ can be neglected, let us solve this equation in the case where $\yin=0$, where it has solution
\beq
\label{mean number e-fold}
\mean{\mathcal{\widetilde{N}}}(\zin) = \ze^2\, {}_2 F_2\left(1,1,\frac{3}{2},2,-\ze^2\right)-\zin^2\, {}_2 F_2\left(1,1,\frac{3}{2},2,-\zin^2\right),
\eeq
with $ {}_2 F_2 $ the generalised hypergeometric function. This leads to the ratio $\ye/y_\uc \sim \yin/y_\uc\ee^{-\langle\widetilde{\mathcal{N}}\rangle(\zin)(\nu_0+3/2)/(\nu_0-3/2)}$ displayed in the left panel of \Fig{fig:ratio time}.

Large values of $\ze$ correspond to large values of $\mu$, \ie to configurations where the noise is subdominant, see \Eq{Langevin}. In that regime, from \Eq{Langevin rotated new variables:z} $z$ decays exponentially with $\widetilde{N}$ and inflation ends after a time $\widetilde{\mathcal{N}}=\ln(\ze/\zin)$. This is why the ratio displayed in \Fig{fig:ratio time} asymptotes $\ye/y_\uc\simeq (\yin/y_\uc)(\zin/\ze)^{(2\nu_0+3)/(2\nu_0-3)}$. In practice, current constraints on the CMB tensor-to-scalar ratio~\cite{Planck:2018jri} impose the upper bound $H<7\times 10^{13}\GeV$ at the time CMB scales exit the Hubble radius. Uphill inflation occurs later, hence at a lower value of $H_0$, and using the double-well value $\nu_0\simeq 4.5$ in \Eq{eq:mu:def} thus gives $\mu\gg 3\times 10^3 \sigma^3$. With $\sigma$ typically of order $0.1$, this means that $\mu$, hence $\ze$, is a large parameter. From \Fig{fig:ratio time}, one concludes that, provided $\zin$ is not too large (\ie provided one starts close enough to the critical trajectory), $y$ plays a negligible role in the end-of-inflation condition, which simply becomes $z=\pm\ze$. In that limit, the dynamics of $z$ fully decouples and the stochastic problem becomes one-dimensional. \\

The absorbing boundary at $z=\pm \ze$ models the fact that, as inflation ends, the noise is turned off (given that modes stop crossing out the Hubble radius). Hence the system is drifted away from the inflating region in phase space, and cannot re-enter it. Now, for sufficiently large $\mu$, the noise becomes subdominant around the end-of-inflation surface anyway. Therefore, as we will now argue, the absorbing boundary can be neglected, since the realisations it prevents from re-entering inflation are so rare that they give negligible contributions to the quantities of physical interest.
More precisely, in the absence of absorbing boundaries the dynamics of $y$ and $z$ decouple, and one has $P(y,\z \vert \widetilde{N},\yin,\zin) = \delta[y-\yin \ee^{- \widetilde{N}(2\nu_0+3)/(2\nu_0-3) }] \G(\z\vert \widetilde{N},\zin)$, where ``NB'' stands for ``no boundaries'' and
\beq
\label{Green PDF}
\G\left(\z\Big\vert \widetilde{N},\zin\right) = \frac{\ee^{-\frac{\left(\z-\zin\ee^{\widetilde{N}}\right)^2}{\ee^{2\widetilde{N}}-1}}}{\sqrt{\pi \left(\ee^{2\widetilde{N}}-1\right)}}\, .
\eeq
This Gaussian distribution is indeed a solution of \Eq{Fokker-Planck PDF} such that $\G(\z\vert 0,\zin)=\delta(\z-\zin)$, but it does not vanish on the end-of-inflation surface.\footnote{In the absence of boundaries, \Eq{Langevin rotated new variables:z} has solution $\z(\widetilde{N})=\zin\ee^{\widetilde{N}}+\Delta z(\widetilde{N})$, where $\Delta z(\widetilde{N}) = \ee^{\widetilde{N}}\int_{0}^{\widetilde{N}} \ee^{-\bar{N}}\xi(\bar{N})\dd\bar{N}$. Since $\Delta z$ is linearly related to the Gaussian noises $\xi(\bar{N})$, it is itself a Gaussian random variable and its first two moments can be computed using $\langle \xi(\bar{N}) \rangle = 0$ and $\langle \xi(\bar{N}_1) \xi(\bar{N}_2)\rangle = \delta(\bar{N}_1-\bar{N}_2)$. One obtains $\langle \Delta z\rangle = 0$ and $\langle \Delta z^2 \rangle = (\ee^{2 \widetilde{N}}-1)/2$, hence the solution~\eqref{Green PDF}.} Removing the boundary condition implies that some realisations in \Eq{Green PDF} resume inflation (by crossing back $\pm \ze)$ while they should not. The occurrence of such an event can be assessed by considering the probability $\Pleft$ that, starting from $ \ze $ at initial time, $ \z < \ze $ at some later time $ \widetilde{N} $ (\ie the system inflates again at time $\widetilde{N}$). Using \Eq{Green PDF}, it is given by
\beq
\label{P boundary}
\Pleft\left(\widetilde{N}\right) = \int_{-\infty}^{\ze}\G\left(\z\Big\vert \widetilde{N},\ze\right)\mathrm{d}\z = \frac{1}{2}\left(1-\erf\left\{\ze\left[\exp\left(\widetilde{N}\right)-1\right]\right\}\right)\, .
\eeq
As a function of $\widetilde{N}$, this varies between $1/2$ and $0$, with a transition that occurs when the argument of the error function is of order one, that is at the characteristic time $\widetilde{N}_\uc = \ln(1+1/\ze)$. This means that, when $\widetilde{N}$ is a few times larger than $\widetilde{N}_\uc$ or more, the probability that inflation (wrongly) resumes becomes negligible. The  characteristic time $\widetilde{N}_\uc$ is compared with the mean number of inflationary \efolds~$\langle\widetilde{\mathcal{N}}\rangle$ in the right panel of \Fig{fig:ratio time}. One can see that, when $\ze$ is large, $\langle\widetilde{\mathcal{N}}\rangle\gg \widetilde{N}_\uc$, hence the return probability can be neglected and \Eq{Green PDF} provides a good approximation to the full solution indeed. In what follows, the results obtained from \Eq{Green PDF} will be compared to full numerical simulations of the Langevin equations~\eqref{Langevin rotated new variables:z}-\eqref{Langevin rotated new variables:y}, and this will confirm the validity of these approximations. 
\subsection{First-passage-time probability}
\begin{figure}[t]
\begin{center}
\includegraphics[width=0.49\textwidth]{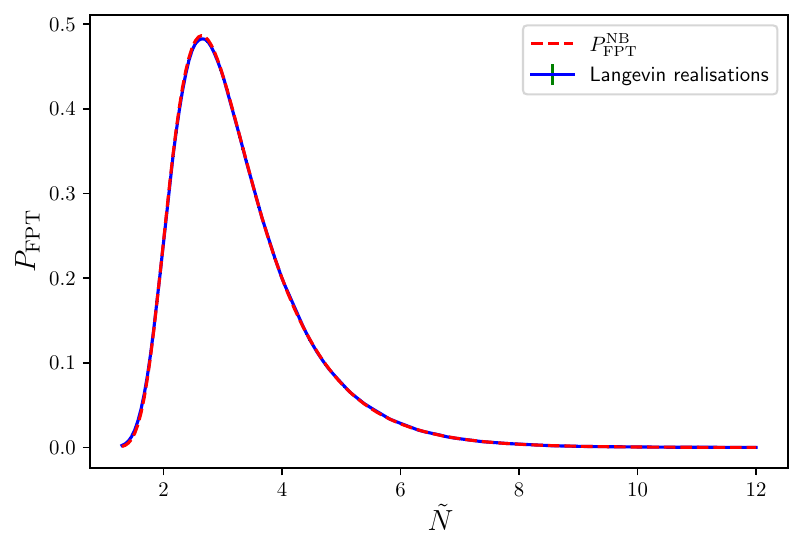}
\includegraphics[width=0.49\textwidth]{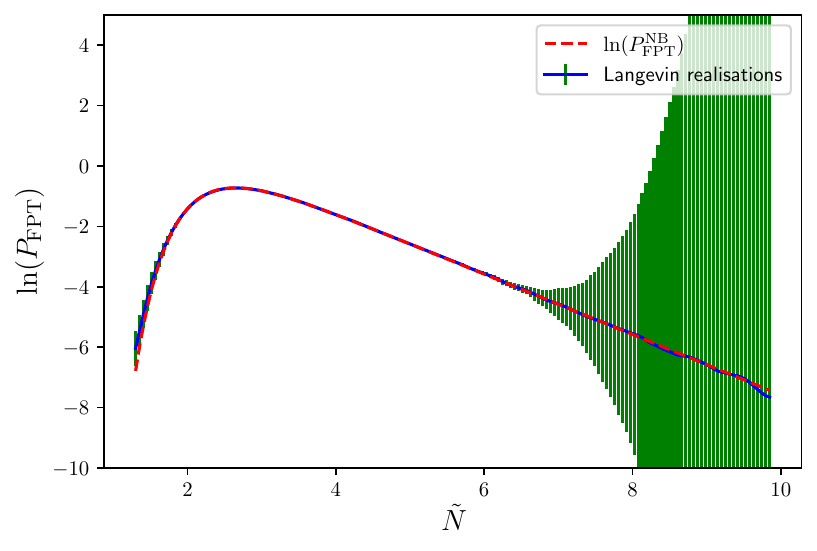}
\caption{First-passage time probability as obtained from the simulation of $ 10^{6} $ realisations of the Langevin equations \eqref{Langevin rotated new variables:z}-\eqref{Langevin rotated new variables:y} (solid blue line), and according to the approximation~\eqref{Green PFPT} (dashed red line). The initial conditions are set at the intersection between the end-of-inflation surface $ \pi = - \pie $ (where inflation resumes) and the $ \y $-axis, see figure \ref{fig:Phase-space}, \ie at $ \zin = 0 $ and $ \yin = \sqrt{2}/(\nu_0+3/2) $. The parameters are set such that $ \sigma = 0.1 $ and $ \ze = 10$. In the stochastic simulations, the error bars (in green) correspond to $ 2\sigma$-estimates of the statistical error using the jackknife method. The right panel shows the distribution in logarithmic scale, to better display the (lower and upper) tails.}
\label{fig:Langevin-pdf-N_simu=10e6-entrÃ©e-inflation_z_end=10}
\end{center}
\end{figure}
The first-passage-time probability can be computed from the solution~\eqref{Green PDF} of the Fokker-Planck equation by introducing the survival probability. This is the probability that, at time $\widetilde{N}$, a given realisation is still inflating. It can be equally written as the probability that the system is still in the inflating region at time $\widetilde{N}$, or as the probability that the first-passage time through the end-of-inflation surface is larger than $\widetilde{N}$. One thus has
\bea
\label{eq:survival:proba}
\int\dd y \int_{-\ze(y)}^{\ze(y)}\dd z \PDF\left(\z,y\Big\vert \widetilde{N},\zin,\yin\right) = \int^\infty_{\widetilde{N}}\PFPT\left(\mathcal{\widetilde{N}}\Big\vert \zin,\yin\right)\mathrm{d}\mathcal{\widetilde{N}} \, .
\eea
By differentiating this expression with respect to $ \widetilde{N} $, and using the Fokker-Planck equation \eqref{Fokker-Planck PDF} where $y$ has been dropped, one obtains
\beq
\label{PFPT function of PDF}
\PFPT(\widetilde{N}\big\vert \zin) = -\int_{-\ze}^{\ze}\widetilde{\mathcal{L}}_{\mathrm{FP}} P(\z\big\vert \widetilde{N},\zin)\mathrm{d}\z
=\left[\left(z - \frac{1}{2}\frac{\partial}{\partial z}\right) P(\z\big\vert \widetilde{N},\zin) \right]_{z=-\ze}^{z=\ze}\, .
\eeq
Inserting \Eq{Green PDF} into this expression yields
\bea
\label{Green PFPT}
    \GFPT(\widetilde{N}\big\vert \zin) =& \frac{2\ze\ee^{2\widetilde{N}}}{\sqrt{\pi \left(\ee^{2\widetilde{N}}-1\right)^3}}
    \exp\left(-\frac{\zin^2\ee^{2\widetilde{N}}+\ze^2}{\ee^{2\widetilde{N}}-1}\right)
    \cosh\left[\frac{\zin\ze }{\sinh(\widetilde{N})}\right] \\
    &\times \left\lbrace 1-\frac{\zin}{\ze}\ee^{-\widetilde{N}}\tanh\left[\frac{\zin\ze}{\sinh(\widetilde{N})}\right]\right\rbrace .
\eea
This formula is compared with numerical simulations of $10^6$ realisations of the Langevin equations~\eqref{Langevin rotated new variables:z}-\eqref{Langevin rotated new variables:y} in \Fig{fig:Langevin-pdf-N_simu=10e6-entrÃ©e-inflation_z_end=10}, where $ \zin = 0 $, $\sigma=0.1$ and $H_0 $ is set such that $ \ze = 10 $. The agreement is excellent, which confirms the validity of the two approximations performed in \Sec{sec:field:proba} (namely dropping $y$ and neglecting the effect of the absorbing boundaries).

\begin{figure}[t]
\begin{center}
\includegraphics[width=0.49\textwidth]{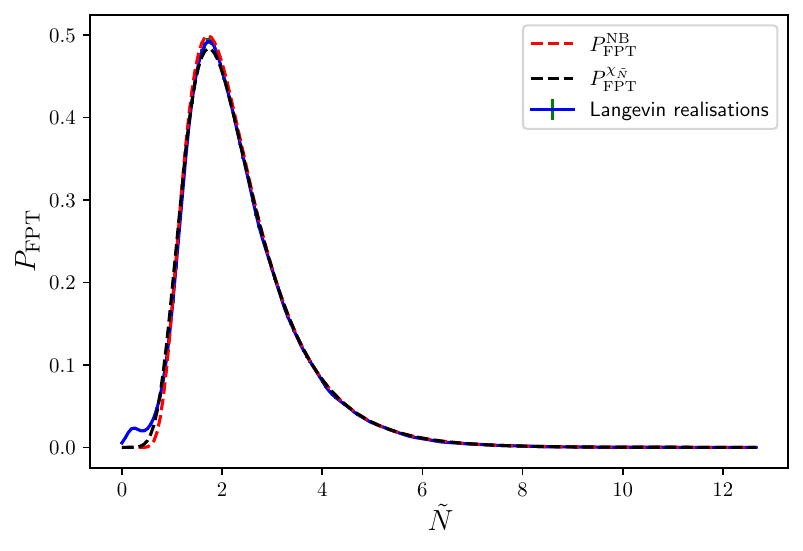}
\includegraphics[width=0.49\textwidth]{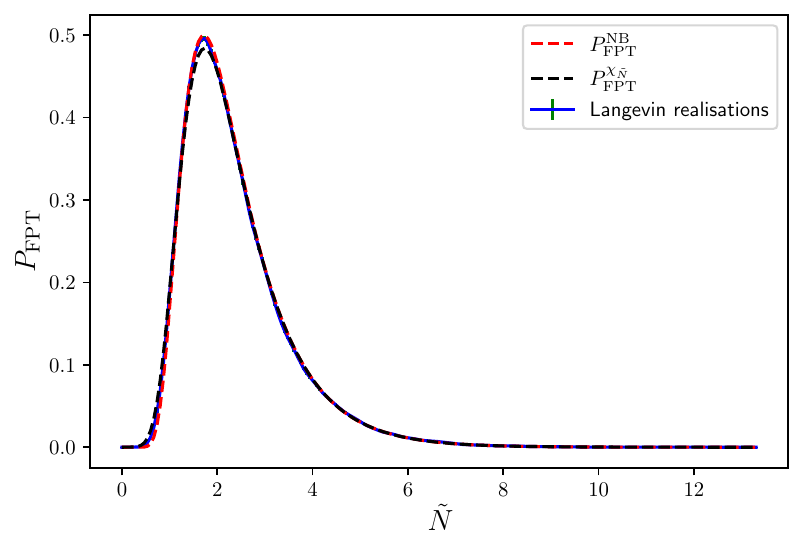}
\caption{Left panel: first-passage time probability in the same situation as in \Fig{fig:Langevin-pdf-N_simu=10e6-entrÃ©e-inflation_z_end=10}, except that now $\ze=4$. In that case, the one-dimensional approximation breaks down at small values of $\widetilde{N}$. Right panel: same as in the left panel, where initial conditions are now taken as $ \yin = \zin = 0 $ (and the one-dimensional approximation becomes exact). 
In both panels, the black lines correspond to the first-passage time probability as reconstructed from the method of poles, see main text and \App{app:FPT}. }
\label{fig:breaking 1d approx}
\end{center}
\end{figure}
For smaller values of $\ze$, the one-dimensional approximation is less reliable, see the discussion around \Fig{fig:ratio time}. This can be seen in \Fig{fig:breaking 1d approx}, which is similar to \Fig{fig:Langevin-pdf-N_simu=10e6-entrÃ©e-inflation_z_end=10} except that $\ze=4$. In that case, at small values of $\widetilde{N}$, the approximation~\eqref{Green PFPT} breaks down, and the full distribution displays a feature that the one-dimensional approximation fails to capture. This can be understood by noting that if inflation is too short, $y$ has not decayed away when it terminates, hence one can no longer discard it.  

Let us also note that, in previous works~\cite{Pattison:2017mbe, Ezquiaga:2019ftu, Pattison:2021oen, Animali:2022otk}, the first-passage time distribution has mostly been obtained by solving the adjoint Fokker-Planck equation~\eqref{eq:adjoint:FP} for the characteristic function. Using the residue theorem, $\PFPT$ can then be written as a sum of decaying exponentials, with exponents that correspond to the poles of the characteristic function. This approach is implemented in details in \App{app:FPT}, where the poles are extracted in the large-$\mu$ (hence large-$\ze$) limit. One obtains an expression that is very similar to \Eq{Green PFPT}, see \Eq{PDF large zend}, which is displayed with the black line in \Fig{fig:breaking 1d approx}. Although it provides a good fit to the full result, the approximation is not as good as \Eq{Green PFPT}, in particular around the maximum of the distribution. This may seem surprising, given that the pole-based approach correctly implements the boundary conditions, contrary to the method employed above. The reason is that, away from the $\widetilde{N}\gg 1$ tail, a large number of poles needs to be summed over, and the individual errors made when approximating each residue in the large-$\ze$ limit accumulate. This is why the method presented here, which consists in neglecting the stochastic noise around the end-of-inflation surface, is in fact better, and in what follows we stick to this approach.

One may object that this method performs well as long as one is interested in a first-passage time through a condition where the stochastic noise is negligible, which is commonly the case around the end-of-inflation surface but not for the generic point $(x_*,y_*)$ which \Eq{eq:zeta:one:point:pdf} integrates over. Although this is true, for Markovian stochastic processes in one dimension (which is the case here), the first-passage time across \emph{any} point can be written in terms of fist-passage times across the end-of-inflation surface only. The reason is that, if $z_*$ lies between $\zin$ and $\ze$, one has $\widetilde{\mathcal{N}}_{\zin\to\ze}=\widetilde{\mathcal{N}}_{\zin\to\z_*}+\widetilde{\mathcal{N}}_{\z_*\to\ze}$ where $\widetilde{\mathcal{N}}_{\zin\to\z_*}$ and $\widetilde{\mathcal{N}}_{\z_*\to\ze}$ are independent variables (due to the Markovian property), hence
\bea
\label{eq:Markov:convolution}
{\PFPT}_{\zin \rightarrow \ze}\left(\widetilde{N}\right) = \int \mathrm{d}\widetilde{N}' {\PFPT}_{\zin \rightarrow \z_*}\left(\widetilde{N}'\right) {\PFPT}_{\z_* \rightarrow \ze}\left(\widetilde{N}-\widetilde{N}'\right)
\eea
for any $ \z_* $. Thus, ${\PFPT}_{\zin \rightarrow \z_*}$ can be expressed in terms of ${\PFPT}_{\zin \rightarrow \ze}$ and ${\PFPT}_{\z_* \rightarrow \ze}$ only, by performing de-convolution of the above expression. This can be done using dedicated de-convolution methods (for a review, see \eg \Refa{2002PASP..114.1051S}); we do not further explore this direction here but that makes this alternative method of broader interest. 

In what follows, ${\PFPT}_{\zin \rightarrow \z_*}$ is rather estimated using the same approximation as above. We start by writing
\bea
\label{eq:survival:proba:generic}
\int_{-\infty}^{z_*} P\left(z\big\vert \widetilde{N},\zin\right) \dd z = p_{\mathrm{exit}}(\zin,z_*) \int_{\widetilde{N}}^\infty {\PFPT}_{\zin \rightarrow \z_*}\left(\widetilde{\mathcal{N}}\right) \dd\widetilde{\mathcal{N}}+1-p_{\mathrm{exit}}(\zin,z_*)
\eea 
as in \Eq{eq:survival:proba}, where we assume $\zin\leq \z_*$ (the case $\zin>z_*$ can be treated along similar lines). In this expression, $p_{\mathrm{exit}}(\zin,z_*)$ corresponds to the probability that, starting from $\zin$, the stochastic process crosses $\z_*$ at least once (which is required for the first-passage time to be defined). Contrary to the situation discussed around \Eq{eq:survival:proba}, where two absorbing boundaries were located at $\pm\ze$ and the exit probability was equal to one, here we have a single absorbing boundary and it is not guaranteed that it gets ever crossed. Formally, $p_{\mathrm{exit}}$ can be related to $P$ by noting that the exit probability is nothing but the probability that, in the asymptotic future, one has $z>z_*$, so
\bea
\label{eq:p:exit:P}
p_{\mathrm{exit}}(\zin,z_*) = 1-\lim_{\widetilde{N}\to\infty}\int_{-\infty}^{z_*} P\left(z\big\vert \widetilde{N},\zin\right)\dd z\, ,
\eea 
which indeed corresponds to \Eq{eq:survival:proba:generic} in the $\widetilde{N}\to\infty$ limit. By differentiating \Eq{eq:survival:proba:generic} with respect to $\widetilde{N}$, one obtains
\bea
\label{eq:PFPT:zin:z:interm}
p_{\mathrm{exit}}(\zin,z_*) {\PFPT}_{\zin \rightarrow \z_*}\left(\widetilde{N}\right) = 
-\int_{-\infty}^{z_*}\widetilde{\mathcal{L}}_{\mathrm{FP}} P(z\big\vert \widetilde{N},\zin)\mathrm{d}z
=\left(z_* - \frac{1}{2}\frac{\partial}{\partial z_*}\right) P(z_*\big\vert \widetilde{N},\zin) \, .
\eea 
Inserting \Eq{Green PDF} into \Eqs{eq:p:exit:P} and~\eqref{eq:PFPT:zin:z:interm} finally leads to $p_{\mathrm{exit}}^{\mathrm{NB}}(\zin,z_*)=[1+\mathrm{erf}(\zin)]/2$ and
\bea 
\label{eq:GFPT:gen}
{\GFPT}_{\zin \rightarrow \z_*>\zin}\left(\widetilde{N}\right) = \frac{2}{1+\mathrm{erf}(\zin)} \frac{z_*\ee^{\widetilde{N}}-\zin}{\sqrt{\pi \left(\ee^{2 \widetilde{N}}-1\right)^3}}
\ee^{\widetilde{N}-\frac{\left(z_*-\zin\ee^{\widetilde{N}}\right)^2}{\ee^{2\widetilde{N}}-1}}\, .
\eea 
Let us stress that this approximation is expected to be reliable only when $z_*$ lies in the drift-dominated region of phase space and not otherwise. One can also check that, when $\zin=0$ and $z_*=\ze$, one recovers \Eq{Green PFPT}, due to the $z\to-z$ symmetry of the problem in that case. 
\section{Backward probability and PBH abundance}
\label{sec:BW and PBH}
Having solved the Fokker-Planck and adjoint Fokker-Planck problems, we are now in a position to evaluate the different terms appearing in \Eq{eq:zeta:one:point:pdf}, and to study phenomenological consequences of that formula such as the probability to form primordial black holes in uphill inflation.
\subsection{Backward probability}
Inserting \Eqs{Green PDF} and~\eqref{Green PFPT} into \Eq{eq:PBW:def}, the backward probability can be approximated by\footnote{The denominator in \Eq{eq:PBW:def} can be computed by integrating \Eq{Green PFPT} over $\widetilde{N}$, but a simpler calculation is to use \Eq{eq:survival:proba} and integrate \Eq{Green PDF} over $z$.}
\bea
\label{Green backward}
\GBW(\z\big\vert\widetilde{N}) = \frac{\sqrt{\pi}}{2}\frac{\GFPT\left(\widetilde{N}\big\vert\z\right)\ee^{z^2}\erfc\left(\lvert \z \rvert\right)}{\erf\left(\frac{\ze}{\sqrt{\ee^{2\widetilde{N}}-1}}\right)}
\eea
where we have set $\zin=0$ (\ie initial conditions are set on the critical trajectory), and where we recall that $\GFPT$ is given by \Eq{Green PFPT}.

\begin{figure}[t]
\begin{center}
\includegraphics[width=0.7\textwidth]{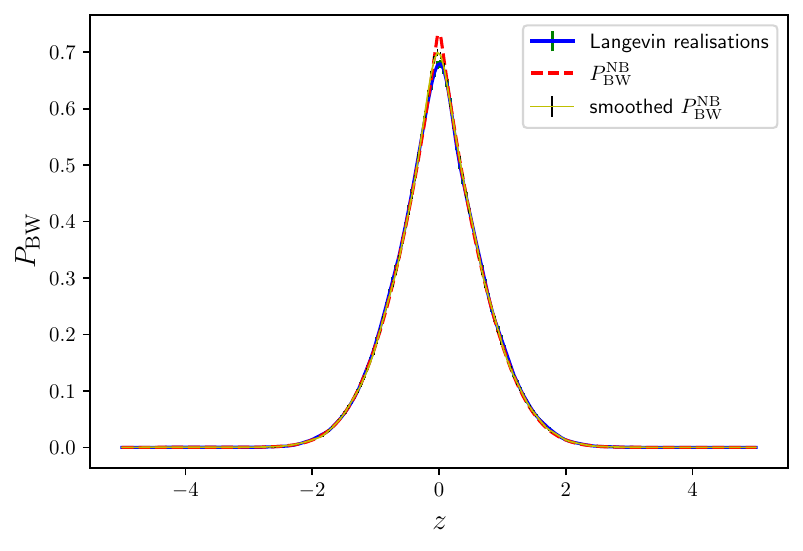}
\caption{Backward probability $\PBW$ of the field value $z$, $N_{\mathrm{BW}}=1$ \efold~before the end of inflation (which corresponds to $\widetilde{N}_{\mathrm{BW}}=\nu_0-3/2$). We have set $ \zin = 0 $ and $ \ze = 10 $. The blue line is reconstructed from $10^5$ realisations of the Langevin equations \eqref{Langevin rotated new variables:z}-\eqref{Langevin rotated new variables:y}, with $2\sigma$-statistical error bars (inferred from jackknife resampling) in green. The red curve corresponds to the no-boundary approximation~\eqref{Green backward}, which is slightly more peaked at the origin. This is partly due to the finite width of the kernel used to reconstruct $\PBW$ from stochastic simulations, so to account for this effect we display in yellow the approximation~\eqref{Green backward} when smoothed with the same kernel.}
\label{fig:PBW}
\end{center}
\end{figure}
This expression is compared with the backward distribution reconstructed from $10^{5}$ simulations of the Langevin equations \eqref{Langevin rotated new variables:z}-\eqref{Langevin rotated new variables:y} in \Fig{fig:PBW}. The agreement is excellent, except close to $\z=0$ where the approximation~\eqref{Green backward} is slightly more peaked. However, when reconstructing $\PBW$ from a finite sample of Langevin realisations, one uses a kernel with a finite width\footnote{In practice, we use the kernel density estimate \texttt{scipy.stats.gaussian\_kde} available in python.}, which has the effect of smoothing out the underlying distribution. To account for this effect, we have smoothed the approximation~\eqref{Green backward} with the same kernel, \ie we have drawn $10^5$ points from that distribution and used the same reconstruction technique as for the Langevin sample. The result is displayed in yellow in \Fig{fig:PBW}, which indeed shows better agreement.
\subsection{Curvature perturbation}
\begin{figure}[t]
\begin{center}
\includegraphics[width=0.49\textwidth]{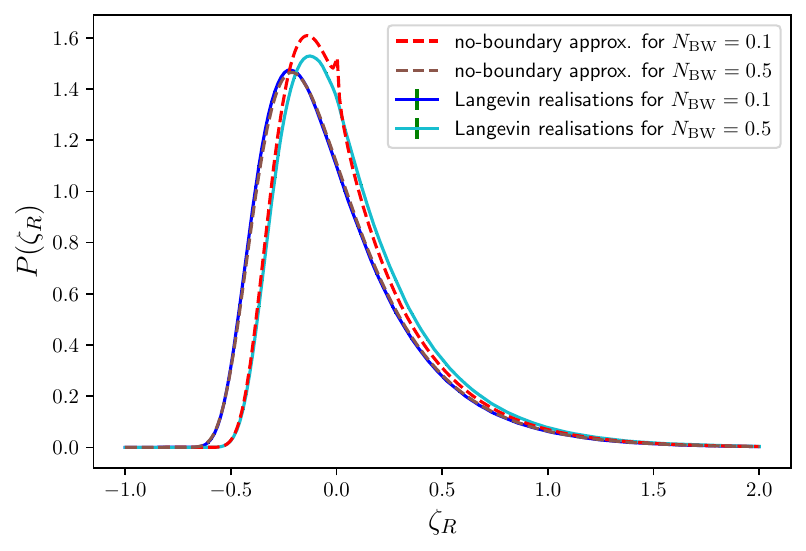}
\includegraphics[width=0.49\textwidth]{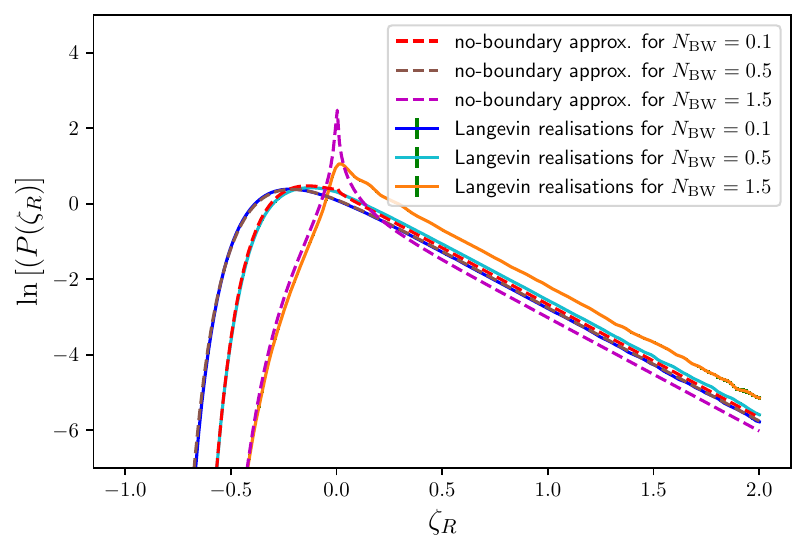}
\caption{One-point distribution function for the curvature perturbation when coarse-grained at the scale $R=(\sigma H_0)^{-1} \ee^{N_{\mathrm{BW}}}$, for a few values of $N_{\mathrm{BW}}$, assuming $\zin=0$ and $\ze=10$. The solid curves are reconstructed from $10^7$ realisations of the Langevin equations~\eqref{Langevin rotated new variables:z}-\eqref{Langevin rotated new variables:y}, with error bars obtained by jackknife resampling. The dashed curves stand for the one-dimensional, no-boundary approximation~\eqref{Curvature perturbation starting zero}, and provide a good fit when $N_{\mathrm{BW}}$ is not too small. The right panel uses a logarithmic scale to better display the tails behaviour (the case $ N_{\mathrm{BW}} = 1 $ is not shown on the left panel to avoid cluttering).}
\label{fig:curvature-perturbation}
\end{center}
\end{figure}

The one-point distribution function of the curvature perturbation, coarse-grained at the scale $R$, is given by \Eq{eq:zeta:one:point:pdf}. In the case $\zin=0$, due to the symmetry $z\to-z$ the integral can be restricted to the range $0\leq \z_*\leq\ze$, and accounting for the rescaling $ \widetilde{N} = (\nu_0-3/2 )N $ one has
\beq
P\left(\zeta_R\right) = \left(2\nu_0-3\right)\!\int_{0}^{\ze}\mathrm{d}\z_*\PBW\left(\z_*\big\vert \widetilde{N}_{\mathrm{BW}}\right){\PFPT}_{0 \rightarrow \z_*}\left[\left(\nu_0-\frac{3}{2}\right)\zeta_R-\langle\mathcal{\widetilde{N}}\rangle\left(\z_*\right)+\langle\mathcal{\widetilde{N}}\rangle\left(\zin\right)\right].
\eeq
Inserting \Eqs{mean number e-fold}, \eqref{eq:GFPT:gen} and~\eqref{Green backward} into that expression, one finds
\beq
\label{Curvature perturbation starting zero}
P^{\mathrm{NB}}\left(\zeta_R\right) = \left(2\nu_0-3\right)\int_{0}^{\ze}\mathrm{d}\z_*\frac{\GFPT\left(\widetilde{N}_{\mathrm{BW}}\big\vert \z_*\right)\ee^{\z_*^2}\erfc\left(\z_*\right)\z_* \alpha\left(\z_*,\zeta_R\right)^2\exp\left[-\frac{\z_*^2}{\alpha\left(\z_*,\zeta_R\right)^2-1}\right]}{\erf\left[\ze \left(\ee^{2\widetilde{N}_{\mathrm{BW}}}-1\right)^{-\frac{1}{2}}\right] \left[\alpha\left(\z_*,\zeta_R\right)^2-1\right]^{\frac{3}{2}}},
\eeq
where $\widetilde{N}_{\mathrm{BW}} = (\nu_0-3/2) N_{\mathrm{BW}}(R) = (\nu_0-3/2)\ln(\sigma H_0 R)$ and we have introduced $ \alpha(\z_*,\zeta_R) \equiv \exp[(\nu_0-3/2)\zeta_R+\z_*^2\, {}_2 F_2(1,1,3/2,2,-\z_*^2)] $. The integral over $\z_*$ needs to be performed numerically, and the result is compared with the distribution reconstructed from $10^7$ realisations of the Langevin equations~\eqref{Langevin rotated new variables:z} and~\eqref{Langevin rotated new variables:y} in \Fig{fig:curvature-perturbation}.
The agreement between the full numerical result and the one-dimensional, no-boundary approximation is good as soon as $N_{\mathrm{BW}}(R)$ is not too large. The reason is that, for large $N_{\mathrm{BW}}$, the backward probability $P_{\mathrm{BW}}(z_*\vert N_{\mathrm{BW}})$ peaks at small values of $\z_*$ (\ie far from the end-of-inflation surface), around which the stochastic dynamics is not drift-dominated. Since the approximation performed in \Eq{eq:GFPT:gen} relied on this condition, it explains why \Eq{Curvature perturbation starting zero} is less reliable at large $N_{\mathrm{BW}}$, especially on the upper tail (see the case $N_{\mathrm{BW}}=1.5$ in the right panel). To solve this issue, one would have to use the deconvolution method alluded to below \Eq{eq:Markov:convolution}. One also recovers the fact that, at large $\zeta_R$, $P(\zeta_R)$ decays exponentially (hence the upper tails are much heavier than in Gaussian statistics).

One may notice that \Eq{Curvature perturbation starting zero} gives a diverging distribution at $ \zeta_R = 0 $, as can be shown analytically by expanding the result around $\zeta_R=0$. This is done in detail in \App{app:diverging density}, where we find that it is of the form $P(\zeta_R)\propto \zeta_R^{-1/2}$. Therefore, although divergent, $P(\zeta_R)$ remains integrable, hence it is well-defined (and so are all its moments). This peaky behaviour can be observed in the right panel, although it is so sharp that it can hardly be resolved (except for $N_{\mathrm{BW}}=1.5$ where it is slightly more pronounced). It also occurs in the distributions sampled by numerical simulations, but it is smoothed away by the reconstruction kernels.

\subsection{PBH abundance}
\begin{figure}[t]
\begin{center}
\includegraphics[width=0.49\textwidth]{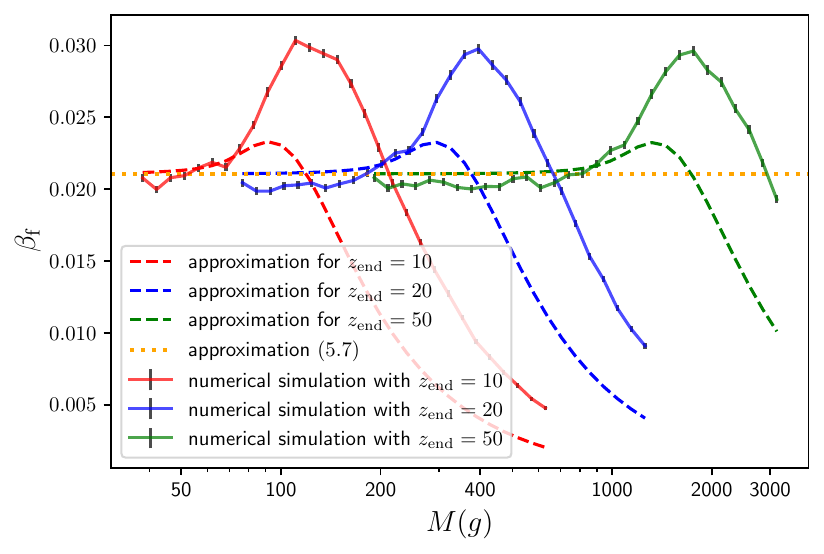}
\includegraphics[width=0.49\textwidth]{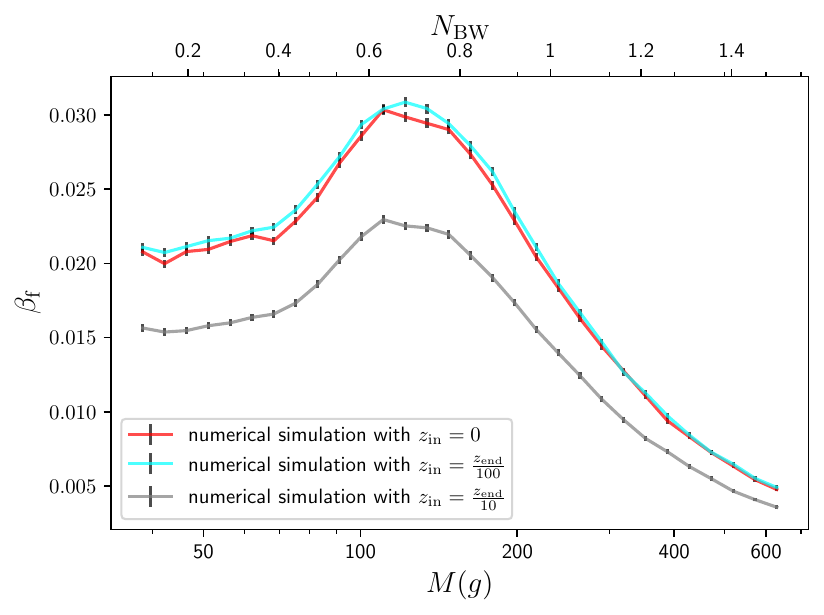}
\caption{Mass fraction of primordial black holes $ \beta_{\mathrm{f}} $ as a function of the mass $ M $ in gram, for $\zeta_\uc=1$. The corresponding value for $ N_{\mathrm{BW}} $ is displayed on the top horizontal axis in the right panel and is related to $M$ via \Eq{eq:mass:NBW} where we have set $\xi=1$ and $\sigma = 0.1$. The value of $ H_0 $ is related to  $ \ze = \mu/\sqrt{\nu_0-3/2} $ using \Eq{eq:mu:def}. On the left panel, $ \beta_{\mathrm{f}} $ is evaluated for different values of $ \ze $ using simulations of the Langevin equations (solid lines, with $95\%$ error bars), compared with the low-mass approximations \eqref{eq:beta:NB} (dashed line) and \eqref{eq:beta:lowmass} (dotted line). On the right panel, $ \beta_{\mathrm{f}} $ is computed via numerical simulations for different values of $ \zin $ and with $ \ze = 10 $.}
\label{fig:mass-fraction}
\end{center}
\end{figure}
One of the phenomenological consequences of the heavy tails observed in $P(\zeta_R)$ is that the probability to form primordial black holes (or other extreme objects such as massive halos~\cite{Ezquiaga:2022qpw}) is enhanced. PBHs form when a large density fluctuation re-enters the Hubble radius and collapses into a black hole. State-of-the-art criteria for PBH formation involve the compaction function~\cite{Shibata:1999zs, Harada:2015yda}, with the mass of the resulting object being assessed via critical scaling relations~\cite{Choptuik:1992jv, Evans:1994pj, Niemeyer:1997mt}. In the present work, for simplicity, we assume that a PBH forms when the local coarse-grained curvature perturbation is larger than a given threshold $\zeta_\uc$ of order one, and assess the PBH abundance with the Press-Schechter estimate
\beq
\label{mass fraction}
\beta_{\mathrm{f}}\left(M\right) = \int_{\zeta_\uc}^{\infty} P\left(\zeta_R\right)\mathrm{d}\zeta_R\int_{N_{\mathrm{BW}}(R)}^\infty  \PFPT(\mathcal{N}\big\vert \zin)  \dd\mathcal{N}
\, .
\eeq
In this expression, the first integral corresponds to the probability that, in regions of the universe inflating at least $N_{\mathrm{BW}}(R)$ \efolds, a given patch of size $R$ collapses into a black hole, while the second integral is the probability that at least $N_{\mathrm{BW}}(R)$ \efolds~are inflated. This second term is usually not considered  since it is close to one in models producing long phases of inflation, but in uphill inflation, which typically lasts for a few \efolds~only, it must be added explicitly.\footnote{Formally, this term cancels out with the denominator of \Eq{eq:PBW:def}, which appears in $P\left(\zeta_R\right)$. When using the algorithm detailed at the end of \Sec{sec:StochaDeltaN}, it simply means that trajectories inflating less than $N_{\mathrm{BW}}(R)$ \efolds~are thrown away, but their number is recorded in order to post-weigh the mass fraction.}

In \Eq{mass fraction}, $M$ corresponds to some fraction $\xi$ of the mass contained in a Hubble patch at the time $R$ re-enters the Hubble radius. Assuming that PBHs form in the radiation era (which is the case for the scales that emerge during the last few \efolds~of inflation if reheating is quick enough), one has
\bea
\label{eq:mass:NBW}
M=\xi \frac{\Mp^2}{\sigma^2 H_0} \ee^{2 N_{\mathrm{BW}}(R)}\, .
\eea 
When $\zin=0$, in the no-boundary approximation developed above one can insert \Eqs{Curvature perturbation starting zero} and~\eqref{Green PFPT} into \Eq{mass fraction} and express the integral over $\zeta_R$ in terms of the error function, leading to
\bea
\label{eq:beta:NB}
\beta_{\mathrm{f}}^{\mathrm{NB}}\left(M\right) =  \int_0^{\ze}\dd z_*\sqrt{\pi}\,\GFPT\left(\widetilde{N}_{\mathrm{BW}}\big\vert \z_*\right)\ee^{\z_*^2}\erfc\left(\z_*\right) 
\erf\left[\frac{\z_* }{\sqrt{\alpha^2\left(\z_*,\zeta_\uc\right)-1}}\right]  ,
\eea 
where we recall that $\GFPT$ is given by \Eq{Green PFPT}.

This expression is compared with the mass fraction extracted from numerical integrations of the Langevin equations~\eqref{Langevin rotated new variables:z}-\eqref{Langevin rotated new variables:y}, following the procedure outlined in \Sec{sec:StochaDeltaN},  in the left panel of \Fig{fig:mass-fraction}. One can check that it gives a good fit at low values of $N_{\mathrm{BW}}$, which is expected since in that case the backward probability peaks at values of $\z_*$ that fall in the drift-dominated region, as explained above. In fact, \Eq{eq:beta:NB} can be further expanded in the limit $\ze \ee^{-\widetilde{N}_{\mathrm{BW}}}\gg 1$, where one obtains
\bea
\label{eq:beta:lowmass}
\beta_{\mathrm{f}}\left(M\right) \simeq  \mathrm{erf}\left[\frac{1}{2}\ee^{-\left(\nu_0-\frac{3}{2}\right)\zeta_\uc-\frac{\gamma_{\mathrm{E}}}{2}}\right]
\eea 
where $\gamma_{\mathrm{E}}\simeq 0.57$ is the Euler-Mascheroni constant. This formula is displayed with the dotted line in the left panel of \Fig{fig:mass-fraction} and correctly captures the mass fraction at low masses. For larger masses it becomes less accurate, but given that the mass fraction does not vary across orders of magnitude in the relevant range of masses, and since our ability to reproduce its detailed shape is anyway limited by the crude Press-Schechter, curvature-perturbation based method we have employed, only the order of magnitude of $\beta_{\mathrm{f}}$ can be safely assessed here. To that end, \Eq{eq:beta:lowmass} provides a reliable estimate.

A striking feature of \Eq{eq:beta:lowmass} is that it neither depends on the mass, nor on $\ze$, \ie on the energy scale at which uphill inflation occurs. Usually, the abundance of black holes increases with the energy density during inflation, since this sets the size of cosmological perturbations. Here, the details of the background dynamics are such that this dependence cancels out, and one obtains mass fractions of order $10^{-2}$ (with $\nu_0\simeq 4.5$ as obtained in the double-well model described in \Sec{sec:double:well}). During the radiation era, the abundance of black holes (which behave as pressure-less matter) grows linearly with the scale factor, hence after a few \efolds~of radiation, PBHs dominate the energy budget of the universe. This conclusion is, again, independent of the energy scale at which hilltop inflation proceeds.
In this scenario, {if inflation proceeds at high-enough energy (namely $ H>10^4\, \GeV$, or equivalently, $\rho^{1/4}> 10^{11} \, \GeV$), PBHs are produced with low masses, smaller than $10^9\, \mathrm{g}$, so they Hawking-evaporate before big-bang nucleosynthesis. This implies that the only imprint left from the PBH-dominated phase is a background of gravitational waves, as studied recently in \Refs{Inomata:2020lmk, Papanikolaou:2020qtd, Domenech:2020ssp}. Otherwise, this scenario is excluded since the universe would remain PBH-dominated at big-bang nucleosynthesis.
\section{Conclusion}
\label{sec:Conclusions}
Let us now summarise our main findings. In this work, we have investigated uphill inflation, where inflation proceeds as the inflaton climbs up a local maximum in its potential energy. If the initial velocity of the inflaton is close to the value classically required to freeze near the local maximum, we have found that inflation proceeds in a new regime, which is neither close to slow roll, fast roll nor ultra-slow roll, and where all three terms in the Klein-Gordon equation are of comparable size. In that regime, although the second and third Hubble-flow functions are singular at the phase-space origin, the Mukhanov-Sasaki effective mass remains regular, and the curvature perturbation $\zeta$ grows as $a^9$, i.e. much faster than in slow roll ($\zeta$ constant) or ultra slow roll ($\zeta\propto a^3$).

This suggests that quantum fluctuations destabilise the freezing of the inflaton at the local maximum by exciting the unstable direction of phase space, and we have studied this mechanism using the stochastic-inflation formalism. Although two-dimensional, the Langevin equations are tractable since only the unstable direction receives a contribution from the noise, while the stable direction remains classical. From the results of \Refa{Tada:2021zzj} we have derived a simple algorithm to extract the statistics of the coarse-grained curvature perturbation $\zeta_R$, which is numerically not more expensive than the computation of a mere first-passage time distribution, and which is fully generic. We have applied this method to the problem at hand and obtained the one-point distribution of $\zeta_R$, as well as the probability for it to exceed the threshold value required to form primordial black holes.

We have also interpreted these results by means of analytical approximations. To that end, we noticed that instead of solving the adjoint Fokker-Planck problem, which requires to impose non-trivial boundary conditions, the first-passage time distributions can be obtained from the solutions of the Fokker-Planck equation directly. The Fokker-Planck equation also comes with absorbing boundary conditions on the end-of-inflation surface, however they can be neglected if one assumes that the stochastic noise is subdominant when inflation ends, which in practice we found to be an excellent approximation. We compared it to the more traditional characteristic-function method, and found that it provides more accurate, and remarkably simpler to derive, results. 

The distribution function of $\zeta_R$ features exponential tails, as is now known to be a ubiquitous consequence of quantum diffusion. The abundance of primordial black holes was found to be independent of the energy scale at which uphill inflation proceeds, and to only involve the curvature of the potential in Hubble units, which is of order unity when uphill inflation is realised in double-well models as explained in \Sec{sec:double:well}. In that case, PBHs are formed with an abundance of order $10^{-2}$, so they quickly dominate the universe content after the first few \efolds~of the radiation era. Being formed with ultra-light masses, they evaporate before big-bang nucleosynthesis (provided inflation proceeds at high-enough energy), but leave a stochastic gravitational-wave background behind~\cite{Inomata:2020lmk, Papanikolaou:2020qtd, Domenech:2020ssp}, possibly as their only imprint. \\

Let us note that, in double-well potentials, the initial velocity required for the field to classically freeze at the local maximum of the potential corresponds to the one inherited from the preceding attractor phase of slow-roll inflation, if the width $\fo$ matches the critical value $\fo^\mathrm{cri}$ given in \Eq{eq:phi0:cri}. This parameter thus needs to be sufficiently fine tuned, which may not come as a surprise since models producing PBHs are  generically known to rely on some fine tuning. Before closing this paper, let us try to better assess the amount of fine tuning that is required.

\begin{figure}[t]
\begin{center}
\includegraphics[width=0.7\textwidth]{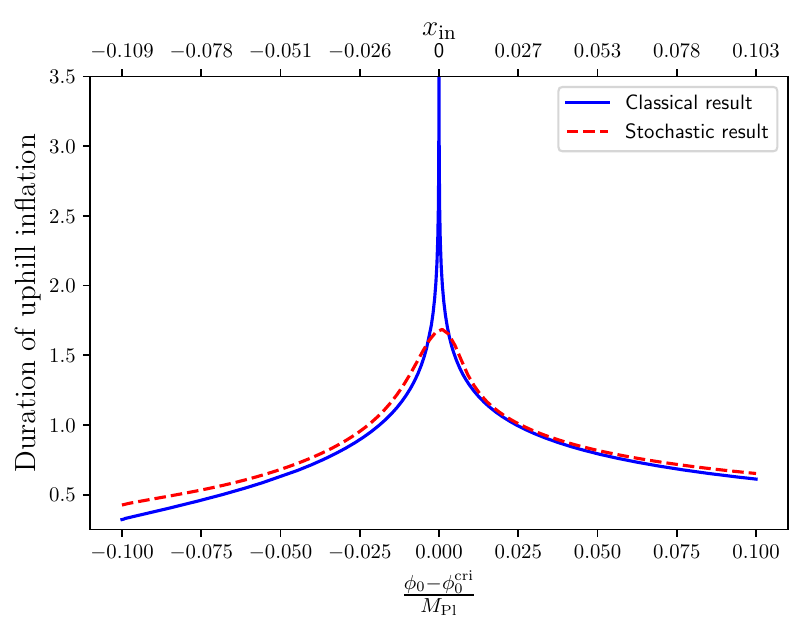}
\caption{Same as in the right panel of \Fig{fig:DWIpot}, \ie~number of inflationary \efolds~realised between the first two crossings of a local minimum in the double-well model discussed in \Sec{sec:double:well}, where the mean number of stochastic \efolds~is also displayed. On the top horizontal axis, the value of $\xin$ that corresponds to $\fo$ is displayed, see main text.}
\label{fig:duration inflation stochastic}
\end{center}
\end{figure}
If $\fo = \fo^\mathrm{cri} $, initial conditions are set on the unstable branch exactly, \ie $\x_\uin=0$. A deviation of $\fo$ from its critical value can thus be described by a non-vanishing value for $\x_\uin$, which simply translates the fact that by tweaking the potential, one may either overshoot the local maximum or turn back before reaching it. In the first case, corresponding to $\fo<\fo^\mathrm{cri}$, the effective value of $x_\uin$ can be identified by matching the field velocity when crossing the potential maximum. In the second case, corresponding to $\fo>\fo^\mathrm{cri}$, it can be done by matching the field value when turning back. This leads to the function $x_\uin(\fo)$ used to label the upper horizontal axis in \Fig{fig:duration inflation stochastic}. There, the mean duration of uphill inflation is displayed, and one can check that the classical divergence observed in \Fig{fig:DWIpot}, and reproduced here with the blue curve for convenience, is renormalised away by quantum diffusion. When $\fo$ deviates more than a few percents from its critical value, the duration of uphill inflation is reduced, and stochastic effects become less important, which explains why the stochastic and classical curves gets closer to each other (the reason why they do not exactly match far from $\fo^\mathrm{cri}$ is because, in the classical problem, the full Klein-Gordon and Friedmann equations are solved, while in the stochastic problem their linearised version around the phase-space origin is employed).

From \Fig{fig:duration inflation stochastic} one can also see that a few-percent deviation for $\fo$ corresponds to working with $\xin$ of order a few percents too, which in turn corresponds to $\zin/\ze$ of order a few percents [using that $\zin/\ze = (\nu_0/2-3/4)\xin$]. This is why, in the right panel of \Fig{fig:mass-fraction}, we show the PBH mass fraction obtained from solving the coupled Langevin equations with two non-vanishing values of $\zin/\ze$, namely $0.01$ and $0.1$. We observe that, although the mass fraction decreases with $\zin/\ze$ as expected, even for $\zin/\ze=0.1$ the abundance of black holes is still of a similar order of magnitude. This corresponds to working with $\xin\simeq 0.07$, hence a value of $\fo$ that is $10\%$ larger than $\fo\equiv\fo^\mathrm{cri} $. As a consequence, the PBH production mechanism seems to be subject to much less fine tuning in uphill inflation than in most alternative models. This makes it an attractive scenario for the production of ultra-light PBHs in the early universe.

\section*{Acknowledgements}
It is a pleasure to thank Chiara Animali, Louis-Pierre Chaintron and Sergey Sibiryakov for valuable discussions.


\appendix

\section{First-passage-time problem with the characteristic-function method}
\label{app:FPT}

The first-passage time distribution $ \PFPT $ satisfies the adjoint Fokker-Planck equation~\eqref{eq:adjoint:FP}, which we reproduce here for simplicity:
\beq
\frac{\partial \PFPT}{\partial \widetilde{N}} = \mathcal{L}_{\mathrm{FP}}^{\dag}\PFPT\, .
\eeq
The boundary conditions for $ \PFPT $ are given by $ \PFPT(\widetilde{N},\pi = \pm \pie) = \delta(\widetilde{N}) $. This is a partial differential equation, the dimensionality of which can be reduced by introducing the characteristic function
\bea
\label{eq:char:def}
\chafunc\left(\widetilde{t}\big\vert z, y\right) = \left\langle \ee^{i \widetilde{t} \widetilde{\mathcal{N}}} \right\rangle = \int_{0}^\infty \dd \widetilde{N} \PFPT\left(\widetilde{N}\big\vert z, y\right)  \ee^{i \widetilde{t} \widetilde{N}}\, .
\eea 
It satisfies the adjoint Fokker-Planck equation
\beq
-i\widetilde{t}\chafunc = \mathcal{L}_{\mathrm{FP}}^{\dag}\chafunc,
\eeq
with boundary conditions $ \chafunc\left(\widetilde{t}\vert \pi = \pm\pie\right) = 1 $. The adjoint Fokker-Planck operator is given in \Eq{eq:adjoint:FP:operator}, so one has to solve
\beq
\label{characteristic function}
-i\widetilde{t}\chafunc = \left[\frac{1}{2}\frac{\partial^2}{\partial \z^2} - \left(\frac{\nu_0+\frac{3}{2}}{\nu_0-\frac{3}{2}}\right)\y\frac{\partial}{\partial \y} + \z\frac{\partial}{\partial \z}\right]\chafunc\, .
\eeq

The structure of the adjoint Fokker-Planck operator allows one to look for solutions that are separable in the variables $ \z $ and $ \y $, \ie that are of the form
\bea
\chafunc\left(\widetilde{t}\big\vert \z,\y\right) = \sum_n \y^n f_n(\z,\widetilde{t}) \, ,
\eea 
where the functions $ f_n $ satisfy
\beq
\label{f_n}
\frac{1}{2}f_n''\left(\z,\widetilde{t}\right) + \z f_n'\left(\z,\widetilde{r}\right) + \left[i\widetilde{t}-\left(\frac{\nu_0+\frac{3}{2}}{\nu_0-\frac{3}{2}}\right)n\right]f_n\left(\z,\widetilde{t}\right) = 0\, .
\eeq
Here, a prime denotes a derivative with respect to $ \z $. Let $ M \left(a,b,\rho\right) $ denote the confluent hypergeometric function of the first kind. Then, the most general solution of \Eq{f_n} is given by
\beq
f_n\left(\z,\widetilde{t}\right) = e^{-z^2}\left\{A_n M\left[a_n\left(\widetilde{t}\right),\frac{1}{2},\z^2\right] + B_n \lvert \z \rvert M\left[a_n\left(\widetilde{t}\right)+\frac{1}{2},\frac{3}{2},\z^2\right]\right\},
\eeq
where $ A_n $ and $ B_n $ are two integration constants and 
\bea
a_n\left(\widetilde{t}\right) = \frac{1}{2}\left(1+\frac{\nu_0+\frac{3}{2}}{\nu_0-\frac{3}{2}}n-i\widetilde{t}\right) \, .
\eea
We want the solution to be regular at $ \z = 0 $ (at least $ f_n' $ should be continuous everywhere), which yields $ B_n = 0$ for all $ n $. This finally leads to
\beq
\label{general solution FP adjoint}
\chafunc\left(\widetilde{t},\z,\y\right) = e^{-z^2}\sum_n A_n \y^n M\left[a_n\left(\widetilde{t}\right),\frac{1}{2},\z^2\right]\, .
\eeq

The next task is to set the constants $A_n$ by using the boundary conditions. This is difficult to do in general, but a simplification arises in the large $\mu$ (or equivalently large $\ze)$ limit, where as argued in \Sec{sec:Fokker Planck}, $\y$ can be dropped and the stochastic problem becomes one-dimensional. In this regime, \Eq{general solution FP adjoint} simply becomes
\beq
\chafunc^{\left(0\right)}\left(\widetilde{t},\z\right) = A_0 e^{-\z^2}M\left[a_0\left(\widetilde{t}\right),\frac{1}{2},\z^2\right]\, .
\eeq
The boundary condition $ \chafunc^{\left(0\right)}\left(\widetilde{t},\z= \pm \ze\right) = 1 $ thus leads to
\beq
\label{chi_N^0}
\chafunc^{\left(0\right)}\left(\widetilde{t},\z\right) = e^{-\left(\z^2-\ze^2\right)}\frac{M\left[a_0\left(\widetilde{t}\right),\frac{1}{2},\z^2\right]}{M\left[a_0\left(\widetilde{t}\right),\frac{1}{2},\ze^2\right]}\, ,
\eeq
where we recall that $ \ze = \mu\pie/\sqrt{2\nu-3}$.

The first-passage-time distribution can then be obtained by noting that \Eq{eq:char:def} is nothing but a Fourier transform, which can be inverted as
\bea
\PFPT(\widetilde{N}\big\vert \z,\y)=\frac{1}{2\pi}\int_{-\infty}^\infty \dd\widetilde{t}\, \chafunc\left(\widetilde{t}\big\vert \z,\y\right)\ee^{-i \widetilde{t} \widetilde{N}}\, .
\eea 
This integral can be evaluated by means of the residue theorem. To this end, one first has to identify the poles of the characteristic function, and compute the residues around those poles. From \Eq{chi_N^0}, the pole equation reads
\beq
M\left[a_0\left(\widetilde{t}\right),\frac{1}{2},\ze^2\right] = 0\, ,
\eeq
the solutions of which are denoted $ \widetilde{t}_i=-i \Lambda_i$, so
\beq
\label{pole equation}
M\left[\frac{1}{2}\left(1-\Lambda_i\right),\frac{1}{2},\ze^2\right] = 0\, .
\eeq

This equation has no analytical solution, but since it was derived under the assumption that $\ze$ is large, it can be further expanded in that limit. To this end, we make use of the formula~\cite{2011ConPh..52..497T}
\beq
\label{large z expansion}
M\left(a,b,\rho\right) = \frac{e^{\rho}\rho^{a-b}\Gamma(b)}{\Gamma(a)}\sum_{s=0}^{\infty}\frac{(1-a)_s(b-a)_s}{s!}\rho^{-s}\, ,
\eeq
where the Pochhammer symbol is defined as $ (d)_s = \prod_{k=0}^{s-1}(d+k) $. When $\rho=\ze$ is large, the leading term dominates the above expansion, which vanishes when $a=(1-\Lambda_i)/2$ is a non-positive integer. This leads to
\bea
\Lambda_i \simeq 1+2i\quad\text{for}\quad i\in\mathbb{N}\, .
\eea
We have checked that this expression provides a good approximation to the first poles obtained from numerically solving \Eq{pole equation}. 

The residues are then obtained as 
\bea
\label{residu large zend}
a_i(z)=-i\left[\frac{\partial}{\partial\widetilde{t}} \chi_{\mathcal{N}}^{-1}\left(\widetilde{t}=-i\Lambda_i\big\vert z\right)\right]^{-1}
= \frac{2\left(-1\right)^n}{\sqrt{\pi}n!}\ze^{2n+1}e^{-\z^2}M\left(-i,\frac{1}{2},\z^2\right)\, .
\eea
By virtue of the residue theorem, the first-passage time distribution is finally given by~\cite{Ezquiaga:2019ftu}
\bea
\PFPTpole\left(\widetilde{N}\big\vert \zin\right)=\sum_i a_i(\zin)\ee^{-\Lambda_i \widetilde{N}}\, ,
\eea
where the superscript  ``$\chafunc$'' highlights the method by which it has been obtained. When replacing $a_i$ and $\Lambda_i$ by the approximated expressions obtained above, it turns out that the sum over $i$ can be performed exactly using the formula~\cite{2011ConPh..52..497T}
\bea 
M\left(-i,\frac{1}{2},\rho\right) = \sum_{j=0}^i \frac{\left(-1\right)^ji!}{\left(i-j\right)!\left(2j\right)!}\left(4\rho\right)^j \, .
\eea
This yields 

\beq
\label{PDF large zend}
    \PFPTpole\left(\widetilde{N}\big\vert \zin\right) = \frac{2\ze}{\sqrt{\pi \left(\ee^{2\widetilde{N}}-1\right)}} \exp\left(-\frac{\zin^2\ee^{2\widetilde{N}}+\ze^2}{\ee^{2\widetilde{N}}-1}\right)
    \cosh\left[\frac{\zin\ze }{\sinh\left(\widetilde{N}\right)}\right]\, .
\eeq
This expression needs to be compared to the approximation found in \Eq{Green PFPT}, from the solution of the Fokker-Planck equation obtained by neglecting the presence of the absorbing boundaries. The two expressions are similar, and only differ by the ratio
\bea
\label{eq:ratio:PDFs}
\frac{ \GFPT\left(\widetilde{N}\big\vert \zin\right)}{\PFPTpole\left(\widetilde{N}\big\vert \zin\right)}=
\frac{\ee^{2 \widetilde{N}}}{\ee^{2\widetilde{N}}-1}  \left\lbrace 1-\frac{\zin}{\ze}\ee^{-\widetilde{N}}\tanh\left[\frac{\zin\ze}{\sinh(\widetilde{N})}\right]\right\rbrace\, .
\eea
The two approximations thus coincide as long as $\ee^{-2\widetilde{N}}\ll 1$, \ie sufficiently far on the tail. The reason is that, away from the tail, a large number of poles contribute to $\PFPTpole$ (while, in the far tail, the dominant contribution is given by the leading pole only), hence the error made when approximating $\Lambda_i$ accumulates. This is why, in \Fig{fig:breaking 1d approx} where \Eqs{Green PFPT} and~\eqref{PDF large zend} are compared, one finds that \Eq{Green PFPT} gives a better fit close to the maximum of $\PFPT$. 
\begin{figure}[t]
\begin{center}
\includegraphics[width=0.8\textwidth]{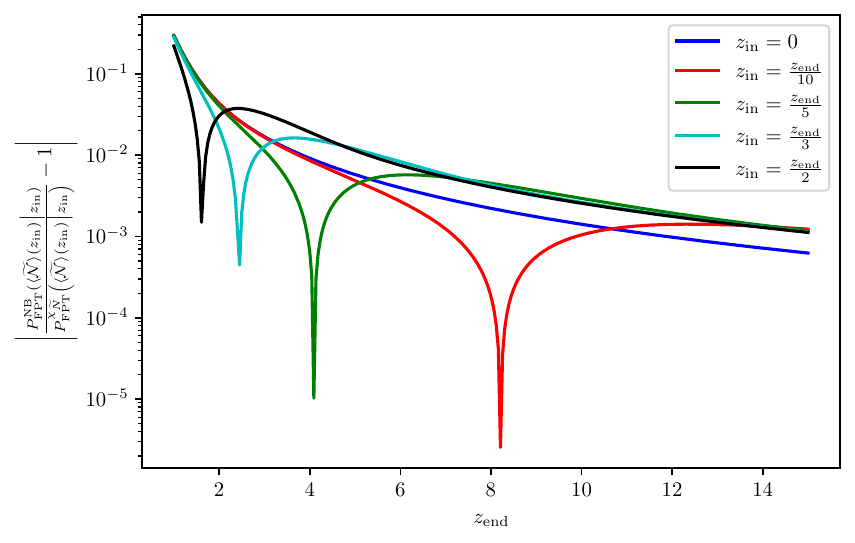}
\caption{Comparison of the two approximations presented in this work to derive $\PFPT$, as given by \Eq{eq:ratio:PDFs}, as a function of $ \ze $ and for a few values of $ \zin $. The first-passage-time distributions are evaluated at $\widetilde{N}=\langle \widetilde{\mathcal{N}} \rangle (\zin)$ given in \Eq{mean number e-fold}.}  
\label{fig:correction}
\end{center}
\end{figure}
More precisely, the ratio~\eqref{eq:ratio:PDFs} is displayed in \Fig{fig:correction} when evaluated at the mean duration $\langle \widetilde{\mathcal{N}} \rangle (\zin)$ given in \Eq{mean number e-fold}. This confirms that, for sufficiently large $\ze$ and sufficiently small $\zin$, the two results agree.
\section{Divergence at $ \zeta_R = 0 $}
\label{app:diverging density}
In this appendix, we study the divergence of the one-point distribution function $P(\zeta_R)$ at $\zeta_R=0$. Let $\epsilon$ denote an infinitesimal small number. We first assume that $\zeta_R>0$. The integral appearing in \Eq{Curvature perturbation starting zero} can be split as
\bea
\label{Curvature perturbation approx positive values}
P^{\mathrm{NB}}\left(\zeta_R\right) =&\frac{\left(2\nu_0-3\right)\GFPT\left(\widetilde{N}_{\mathrm{BW}}\vert 0\right)}{\erf\left\{\ze \left(\ee^{2\widetilde{N}_{\mathrm{BW}}}-1\right)^{-\frac{1}{2}}\right\}}\int_{0}^{\epsilon}\mathrm{d}\z_*\frac{\alpha^2\left(\z_*,\zeta_R\right)\z_*\exp\left[-\frac{\z_*^2}{\alpha\left(\z_*,\zeta_R\right)^2-1}\right]}{\left[\alpha\left(\z_*,\zeta_R\right)^2-1\right]^{\frac{3}{2}}}\\
&+
\left(2\nu_0-3\right)\int_{\epsilon}^{\ze}\mathrm{d}\z_*\frac{\GFPT\left(\widetilde{N}_{\mathrm{BW}}\big\vert \z_*\right)\ee^{\z_*^2}\erfc\left(\z_*\right)\z_* \alpha^2\left(\z_*,\zeta_R\right)\exp\left[-\frac{\z_*^2}{\alpha\left(\z_*,\zeta_R\right)^2-1}\right]}{\erf\left\{\ze \left(\ee^{2\widetilde{N}_{\mathrm{BW}}}-1\right)^{-\frac{1}{2}}\right\} \left[\alpha\left(\z_*,\zeta_R\right)^2-1\right]^{\frac{3}{2}}}
\eea
where in the first line, we have set $\z_*=0$ in the non-diverging prefactor, since keeping $\z_*$ would only bring $\epsilon-$suppressed corrections. We also recall that $ \alpha(\z_*,\zeta_R)= \exp[(\nu_0-3/2)\zeta_R+\z_*^2\, {}_2 F_2(1,1,3/2,2,-\z_*^2)] $. The term responsible for the divergence at small $\zeta_R$ is the integral in the first line, namely
\bea
\label{eq:Iepsilon:def}
I_{\epsilon}\left(\zeta_R\right) \equiv \int_{0}^{\epsilon}\mathrm{d}\z_*\frac{\alpha^2\left(\z_*,\zeta_R\right)\z_*\exp\left[-\frac{\z_*^2}{\alpha\left(\z_*,\zeta_R\right)^2-1}\right]}{\left[\alpha\left(\z_*,\zeta_R\right)^2-1\right]^{\frac{3}{2}}}\, .
\eea
Denoting $\gamma\equiv \exp[(2\nu_0-3)\zeta_R]$, and expanding \Eq{eq:Iepsilon:def} in the limit where both $\epsilon$ and $\gamma-1$ are small (but $\epsilon$ is not necessarily small compared to $ \sqrt{1-\gamma} $), one has
\beq
I_{\epsilon}\left(\zeta_R\right) \simeq \frac{1-\ee^{-\frac{\epsilon^2}{\gamma-1}}}{2\sqrt{\gamma-1}}\, .
\eeq
For $\zeta_R>0$, $\gamma>1$ and $I_\epsilon(\zeta_R)\propto\zeta_R^{-1/2}$. A similar calculation shows that this is also the case for $\zeta_R<0$.
\bibliographystyle{JHEP}
\bibliography{biblio}

\providecommand{\href}[2]{#2}\begingroup\raggedright\begin{thebibliography}{10}

\bibitem{Hawking:1971ei}
S.~Hawking, \emph{{Gravitationally collapsed objects of very low mass}},
  {\emph{Mon. Not. Roy. Astron. Soc.} {\bf 152} (1971) 75}.

\bibitem{Carr:1974nx}
B.~J. Carr and S.~W. Hawking, \emph{{Black holes in the early Universe}},
  {\emph{Mon. Not. Roy. Astron. Soc.} {\bf 168} (1974) 399--415}.

\bibitem{Carr:1975qj}
B.~J. Carr, \emph{{The Primordial black hole mass spectrum}},
  \href{http://dx.doi.org/10.1086/153853}{\emph{Astrophys. J.} {\bf 201} (1975)
  1--19}.

\bibitem{Escriva:2022duf}
A.~Escriv\`a, F.~Kuhnel and Y.~Tada, \emph{{Primordial Black Holes}},
  \href{http://arxiv.org/abs/2211.05767}{{\tt 2211.05767}}.

\bibitem{Starobinsky:1979ty}
A.~A. Starobinsky, \emph{{Spectrum of relict gravitational radiation and the
  early state of the universe}}, {\emph{JETP Lett.} {\bf 30} (1979) 682--685}.

\bibitem{Mukhanov:1981xt}
V.~F. Mukhanov and G.~V. Chibisov, \emph{{Quantum Fluctuations and a
  Nonsingular Universe}}, {\emph{JETP Lett.} {\bf 33} (1981) 532--535}.

\bibitem{Starobinsky:1982ee}
A.~A. Starobinsky, \emph{{Dynamics of Phase Transition in the New Inflationary
  Universe Scenario and Generation of Perturbations}},
  \href{http://dx.doi.org/10.1016/0370-2693(82)90541-X}{\emph{Phys. Lett. B}
  {\bf 117} (1982) 175--178}.

\bibitem{Guth:1982ec}
A.~H. Guth and S.~Pi, \emph{{Fluctuations in the New Inflationary Universe}},
  \href{http://dx.doi.org/10.1103/PhysRevLett.49.1110}{\emph{Phys. Rev. Lett.}
  {\bf 49} (1982) 1110--1113}.

\bibitem{Bardeen:1983qw}
J.~M. Bardeen, P.~J. Steinhardt and M.~S. Turner, \emph{{Spontaneous Creation
  of Almost Scale - Free Density Perturbations in an Inflationary Universe}},
  \href{http://dx.doi.org/10.1103/PhysRevD.28.679}{\emph{Phys. Rev. D} {\bf 28}
  (1983) 679}.

\bibitem{Planck:2018vyg}
{\scshape Planck} collaboration, N.~Aghanim et~al., \emph{{Planck 2018 results.
  VI. Cosmological parameters}},
  \href{http://dx.doi.org/10.1051/0004-6361/201833910}{\emph{Astron.
  Astrophys.} {\bf 641} (2020) A6},
  [\href{http://arxiv.org/abs/1807.06209}{{\tt 1807.06209}}].

\bibitem{Planck:2018nkj}
{\scshape Planck} collaboration, N.~Aghanim et~al., \emph{{Planck 2018 results.
  I. Overview and the cosmological legacy of Planck}},
  \href{http://dx.doi.org/10.1051/0004-6361/201833880}{\emph{Astron.
  Astrophys.} {\bf 641} (2020) A1},
  [\href{http://arxiv.org/abs/1807.06205}{{\tt 1807.06205}}].

\bibitem{SDSS:2005xqv}
{\scshape SDSS} collaboration, D.~J. Eisenstein et~al., \emph{{Detection of the
  Baryon Acoustic Peak in the Large-Scale Correlation Function of SDSS Luminous
  Red Galaxies}}, \href{http://dx.doi.org/10.1086/466512}{\emph{Astrophys. J.}
  {\bf 633} (2005) 560--574},
  [\href{http://arxiv.org/abs/astro-ph/0501171}{{\tt astro-ph/0501171}}].

\bibitem{BOSS:2014hwf}
{\scshape BOSS} collaboration, T.~Delubac et~al., \emph{{Baryon acoustic
  oscillations in the Ly\ensuremath{\alpha} forest of BOSS DR11 quasars}},
  \href{http://dx.doi.org/10.1051/0004-6361/201423969}{\emph{Astron.
  Astrophys.} {\bf 574} (2015) A59},
  [\href{http://arxiv.org/abs/1404.1801}{{\tt 1404.1801}}].

\bibitem{Amendola:2016saw}
L.~Amendola et~al., \emph{{Cosmology and fundamental physics with the Euclid
  satellite}}, \href{http://dx.doi.org/10.1007/s41114-017-0010-3}{\emph{Living
  Rev. Rel.} {\bf 21} (2018) 2}, [\href{http://arxiv.org/abs/1606.00180}{{\tt
  1606.00180}}].

\bibitem{DES:2022qpf}
{\scshape DES} collaboration, C.~Doux et~al., \emph{{Dark energy survey year 3
  results: cosmological constraints from the analysis of cosmic shear in
  harmonic space}}, \href{http://dx.doi.org/10.1093/mnras/stac1826}{\emph{Mon.
  Not. Roy. Astron. Soc.} {\bf 515} (2022) 1942--1972},
  [\href{http://arxiv.org/abs/2203.07128}{{\tt 2203.07128}}].

\bibitem{Martin:2013tda}
J.~Martin, C.~Ringeval and V.~Vennin, \emph{{Encyclop\ae{}dia Inflationaris}},
  \href{http://dx.doi.org/10.1016/j.dark.2014.01.003}{\emph{Phys. Dark Univ.}
  {\bf 5-6} (2014) 75--235}, [\href{http://arxiv.org/abs/1303.3787}{{\tt
  1303.3787}}].

\bibitem{Planck:2018jri}
{\scshape Planck} collaboration, Y.~Akrami et~al., \emph{{Planck 2018 results.
  X. Constraints on inflation}},
  \href{http://dx.doi.org/10.1051/0004-6361/201833887}{\emph{Astron.
  Astrophys.} {\bf 641} (2020) A10},
  [\href{http://arxiv.org/abs/1807.06211}{{\tt 1807.06211}}].

\bibitem{Choudhury:2013woa}
S.~Choudhury and A.~Mazumdar, \emph{{Primordial blackholes and gravitational
  waves for an inflection-point model of inflation}},
  \href{http://dx.doi.org/10.1016/j.physletb.2014.04.050}{\emph{Phys. Lett. B}
  {\bf 733} (2014) 270--275}, [\href{http://arxiv.org/abs/1307.5119}{{\tt
  1307.5119}}].

\bibitem{Kawasaki:2016pql}
M.~Kawasaki, A.~Kusenko, Y.~Tada and T.~T. Yanagida, \emph{{Primordial black
  holes as dark matter in supergravity inflation models}},
  \href{http://dx.doi.org/10.1103/PhysRevD.94.083523}{\emph{Phys. Rev. D} {\bf
  94} (2016) 083523}, [\href{http://arxiv.org/abs/1606.07631}{{\tt
  1606.07631}}].

\bibitem{Garcia-Bellido:2017mdw}
J.~Garcia-Bellido and E.~Ruiz~Morales, \emph{{Primordial black holes from
  single field models of inflation}},
  \href{http://dx.doi.org/10.1016/j.dark.2017.09.007}{\emph{Phys. Dark Univ.}
  {\bf 18} (2017) 47--54}, [\href{http://arxiv.org/abs/1702.03901}{{\tt
  1702.03901}}].

\bibitem{Ezquiaga:2017fvi}
J.~M. Ezquiaga, J.~Garcia-Bellido and E.~Ruiz~Morales, \emph{{Primordial Black
  Hole production in Critical Higgs Inflation}},
  \href{http://dx.doi.org/10.1016/j.physletb.2017.11.039}{\emph{Phys. Lett. B}
  {\bf 776} (2018) 345--349}, [\href{http://arxiv.org/abs/1705.04861}{{\tt
  1705.04861}}].

\bibitem{Germani:2017bcs}
C.~Germani and T.~Prokopec, \emph{{On primordial black holes from an inflection
  point}}, \href{http://dx.doi.org/10.1016/j.dark.2017.09.001}{\emph{Phys. Dark
  Univ.} {\bf 18} (2017) 6--10}, [\href{http://arxiv.org/abs/1706.04226}{{\tt
  1706.04226}}].

\bibitem{Ezquiaga:2018gbw}
J.~M. Ezquiaga and J.~Garc\'\i{}a-Bellido, \emph{{Quantum diffusion beyond
  slow-roll: implications for primordial black-hole production}},
  \href{http://dx.doi.org/10.1088/1475-7516/2018/08/018}{\emph{JCAP} {\bf 08}
  (2018) 018}, [\href{http://arxiv.org/abs/1805.06731}{{\tt 1805.06731}}].

\bibitem{Bhaumik:2019tvl}
N.~Bhaumik and R.~K. Jain, \emph{{Primordial black holes dark matter from
  inflection point models of inflation and the effects of reheating}},
  \href{http://dx.doi.org/10.1088/1475-7516/2020/01/037}{\emph{JCAP} {\bf 01}
  (2020) 037}, [\href{http://arxiv.org/abs/1907.04125}{{\tt 1907.04125}}].

\bibitem{Figueroa:2020jkf}
D.~G. Figueroa, S.~Raatikainen, S.~Rasanen and E.~Tomberg, \emph{{Non-Gaussian
  Tail of the Curvature Perturbation in Stochastic Ultraslow-Roll Inflation:
  Implications for Primordial Black Hole Production}},
  \href{http://dx.doi.org/10.1103/PhysRevLett.127.101302}{\emph{Phys. Rev.
  Lett.} {\bf 127} (2021) 101302}, [\href{http://arxiv.org/abs/2012.06551}{{\tt
  2012.06551}}].

\bibitem{Dvali:2003vv}
G.~Dvali and S.~Kachru, \emph{{New old inflation}},  in \emph{{From Fields to
  Strings: Circumnavigating Theoretical Physics: A Conference in Tribute to Ian
  Kogan}}, pp.~1131--1155, 9, 2003.
\newblock \href{http://arxiv.org/abs/hep-th/0309095}{{\tt hep-th/0309095}}.

\bibitem{Hamada:2014wna}
Y.~Hamada, H.~Kawai, K.-y. Oda and S.~C. Park, \emph{{Higgs inflation from
  Standard Model criticality}},
  \href{http://dx.doi.org/10.1103/PhysRevD.91.053008}{\emph{Phys. Rev. D} {\bf
  91} (2015) 053008}, [\href{http://arxiv.org/abs/1408.4864}{{\tt 1408.4864}}].

\bibitem{Bezrukov:2014bra}
F.~Bezrukov and M.~Shaposhnikov, \emph{{Higgs inflation at the critical
  point}}, \href{http://dx.doi.org/10.1016/j.physletb.2014.05.074}{\emph{Phys.
  Lett. B} {\bf 734} (2014) 249--254},
  [\href{http://arxiv.org/abs/1403.6078}{{\tt 1403.6078}}].

\bibitem{Kitajima:2019ibn}
N.~Kitajima, Y.~Tada and F.~Takahashi, \emph{{Stochastic inflation with an
  extremely large number of $e$-folds}},
  \href{http://dx.doi.org/10.1016/j.physletb.2019.135097}{\emph{Phys. Lett. B}
  {\bf 800} (2020) 135097}, [\href{http://arxiv.org/abs/1908.08694}{{\tt
  1908.08694}}].

\bibitem{Geller:2022nkr}
S.~R. Geller, W.~Qin, E.~McDonough and D.~I. Kaiser, \emph{{Primordial black
  holes from multifield inflation with nonminimal couplings}},
  \href{http://dx.doi.org/10.1103/PhysRevD.106.063535}{\emph{Phys. Rev. D} {\bf
  106} (2022) 063535}, [\href{http://arxiv.org/abs/2205.04471}{{\tt
  2205.04471}}].

\bibitem{Gu:2022pbo}
B.-M. Gu, F.-W. Shu, K.~Yang and Y.-P. Zhang, \emph{{Primordial black holes
  from an inflationary potential valley}},
  \href{http://dx.doi.org/10.1103/PhysRevD.107.023519}{\emph{Phys. Rev. D} {\bf
  107} (2023) 023519}, [\href{http://arxiv.org/abs/2207.09968}{{\tt
  2207.09968}}].

\bibitem{Animali:2022otk}
C.~Animali and V.~Vennin, \emph{{Primordial black holes from stochastic
  tunnelling}},  \href{http://arxiv.org/abs/2210.03812}{{\tt 2210.03812}}.

\bibitem{Atal:2019cdz}
V.~Atal, J.~Garriga and A.~Marcos-Caballero, \emph{{Primordial black hole
  formation with non-Gaussian curvature perturbations}},
  \href{http://dx.doi.org/10.1088/1475-7516/2019/09/073}{\emph{JCAP} {\bf 09}
  (2019) 073}, [\href{http://arxiv.org/abs/1905.13202}{{\tt 1905.13202}}].

\bibitem{Atal:2019erb}
V.~Atal, J.~Cid, A.~Escriv\`a and J.~Garriga, \emph{{PBH in single field
  inflation: the effect of shape dispersion and non-Gaussianities}},
  \href{http://dx.doi.org/10.1088/1475-7516/2020/05/022}{\emph{JCAP} {\bf 05}
  (2020) 022}, [\href{http://arxiv.org/abs/1908.11357}{{\tt 1908.11357}}].

\bibitem{Inomata:2021tpx}
K.~Inomata, E.~McDonough and W.~Hu, \emph{{Amplification of primordial
  perturbations from the rise or fall of the inflaton}},
  \href{http://dx.doi.org/10.1088/1475-7516/2022/02/031}{\emph{JCAP} {\bf 02}
  (2022) 031}, [\href{http://arxiv.org/abs/2110.14641}{{\tt 2110.14641}}].

\bibitem{Cai:2021zsp}
Y.-F. Cai, X.-H. Ma, M.~Sasaki, D.-G. Wang and Z.~Zhou, \emph{{One small step
  for an inflaton, one giant leap for inflation: A novel non-Gaussian tail and
  primordial black holes}},
  \href{http://dx.doi.org/10.1016/j.physletb.2022.137461}{\emph{Phys. Lett. B}
  {\bf 834} (2022) 137461}, [\href{http://arxiv.org/abs/2112.13836}{{\tt
  2112.13836}}].

\bibitem{Cai:2022erk}
Y.-F. Cai, X.-H. Ma, M.~Sasaki, D.-G. Wang and Z.~Zhou, \emph{{Highly
  non-Gaussian tails and primordial black holes from single-field inflation}},
  \href{http://dx.doi.org/10.1088/1475-7516/2022/12/034}{\emph{JCAP} {\bf 12}
  (2022) 034}, [\href{http://arxiv.org/abs/2207.11910}{{\tt 2207.11910}}].

\bibitem{Pi:2022ysn}
S.~Pi and M.~Sasaki, \emph{{Logarithmic Duality of the Curvature
  Perturbation}},  \href{http://arxiv.org/abs/2211.13932}{{\tt 2211.13932}}.

\bibitem{Chen:2013aj}
X.~Chen, H.~Firouzjahi, M.~H. Namjoo and M.~Sasaki, \emph{{A Single Field
  Inflation Model with Large Local Non-Gaussianity}},
  \href{http://dx.doi.org/10.1209/0295-5075/102/59001}{\emph{EPL} {\bf 102}
  (2013) 59001}, [\href{http://arxiv.org/abs/1301.5699}{{\tt 1301.5699}}].

\bibitem{Chen:2013eea}
X.~Chen, H.~Firouzjahi, E.~Komatsu, M.~H. Namjoo and M.~Sasaki, \emph{{In-in
  and $\delta N$ calculations of the bispectrum from non-attractor single-field
  inflation}},
  \href{http://dx.doi.org/10.1088/1475-7516/2013/12/039}{\emph{JCAP} {\bf 12}
  (2013) 039}, [\href{http://arxiv.org/abs/1308.5341}{{\tt 1308.5341}}].

\bibitem{Yokoyama:1998pt}
J.~Yokoyama, \emph{{Chaotic new inflation and formation of primordial black
  holes}}, \href{http://dx.doi.org/10.1103/PhysRevD.58.083510}{\emph{Phys. Rev.
  D} {\bf 58} (1998) 083510},
  [\href{http://arxiv.org/abs/astro-ph/9802357}{{\tt astro-ph/9802357}}].

\bibitem{Pattison:2017mbe}
C.~Pattison, V.~Vennin, H.~Assadullahi and D.~Wands, \emph{{Quantum diffusion
  during inflation and primordial black holes}},
  \href{http://dx.doi.org/10.1088/1475-7516/2017/10/046}{\emph{JCAP} {\bf 10}
  (2017) 046}, [\href{http://arxiv.org/abs/1707.00537}{{\tt 1707.00537}}].

\bibitem{Ezquiaga:2019ftu}
J.~M. Ezquiaga, J.~Garc\'\i{}a-Bellido and V.~Vennin, \emph{{The exponential
  tail of inflationary fluctuations: consequences for primordial black holes}},
  \href{http://dx.doi.org/10.1088/1475-7516/2020/03/029}{\emph{JCAP} {\bf 03}
  (2020) 029}, [\href{http://arxiv.org/abs/1912.05399}{{\tt 1912.05399}}].

\bibitem{Panagopoulos:2019ail}
G.~Panagopoulos and E.~Silverstein, \emph{{Primordial Black Holes from
  non-Gaussian tails}},  \href{http://arxiv.org/abs/1906.02827}{{\tt
  1906.02827}}.

\bibitem{Pattison:2021oen}
C.~Pattison, V.~Vennin, D.~Wands and H.~Assadullahi, \emph{{Ultra-slow-roll
  inflation with quantum diffusion}},
  \href{http://dx.doi.org/10.1088/1475-7516/2021/04/080}{\emph{JCAP} {\bf 04}
  (2021) 080}, [\href{http://arxiv.org/abs/2101.05741}{{\tt 2101.05741}}].

\bibitem{Achucarro:2021pdh}
A.~Achucarro, S.~Cespedes, A.-C. Davis and G.~A. Palma, \emph{{The hand-made
  tail: non-perturbative tails from multifield inflation}},
  \href{http://dx.doi.org/10.1007/JHEP05(2022)052}{\emph{JHEP} {\bf 05} (2022)
  052}, [\href{http://arxiv.org/abs/2112.14712}{{\tt 2112.14712}}].

\bibitem{Hooshangi:2021ubn}
S.~Hooshangi, M.~H. Namjoo and M.~Noorbala, \emph{{Rare events are
  nonperturbative: Primordial black holes from heavy-tailed distributions}},
  \href{http://dx.doi.org/10.1016/j.physletb.2022.137400}{\emph{Phys. Lett. B}
  {\bf 834} (2022) 137400}, [\href{http://arxiv.org/abs/2112.04520}{{\tt
  2112.04520}}].

\bibitem{Ezquiaga:2022qpw}
J.~M. Ezquiaga, J.~Garc\'\i{}a-Bellido and V.~Vennin, \emph{{Could ''El Gordo''
  be hinting at primordial quantum diffusion?}},
  \href{http://arxiv.org/abs/2207.06317}{{\tt 2207.06317}}.

\bibitem{Cohen:2022clv}
T.~Cohen, D.~Green and A.~Premkumar, \emph{{Large Deviations in the Early
  Universe}},  \href{http://arxiv.org/abs/2212.02535}{{\tt 2212.02535}}.

\bibitem{Starobinsky:1986fx}
A.~A. Starobinsky, \emph{{Stochastic De Sitter (inflationary) stage in the
  early universe}},
  \href{http://dx.doi.org/10.1007/3-540-16452-9_6}{\emph{Lect. Notes Phys.}
  {\bf 246} (1986) 107--126}.

\bibitem{Fujita:2013cna}
T.~Fujita, M.~Kawasaki, Y.~Tada and T.~Takesako, \emph{{A new algorithm for
  calculating the curvature perturbations in stochastic inflation}},
  \href{http://dx.doi.org/10.1088/1475-7516/2013/12/036}{\emph{JCAP} {\bf 12}
  (2013) 036}, [\href{http://arxiv.org/abs/1308.4754}{{\tt 1308.4754}}].

\bibitem{Vennin:2015hra}
V.~Vennin and A.~A. Starobinsky, \emph{{Correlation Functions in Stochastic
  Inflation}},
  \href{http://dx.doi.org/10.1140/epjc/s10052-015-3643-y}{\emph{Eur. Phys. J.
  C} {\bf 75} (2015) 413}, [\href{http://arxiv.org/abs/1506.04732}{{\tt
  1506.04732}}].

\bibitem{Tada:2021zzj}
Y.~Tada and V.~Vennin, \emph{{Statistics of coarse-grained cosmological fields
  in stochastic inflation}},
  \href{http://dx.doi.org/10.1088/1475-7516/2022/02/021}{\emph{JCAP} {\bf 02}
  (2022) 021}, [\href{http://arxiv.org/abs/2111.15280}{{\tt 2111.15280}}].

\bibitem{Chowdhury:2019otk}
D.~Chowdhury, J.~Martin, C.~Ringeval and V.~Vennin, \emph{{Assessing the
  scientific status of inflation after Planck}},
  \href{http://dx.doi.org/10.1103/PhysRevD.100.083537}{\emph{Phys. Rev. D} {\bf
  100} (2019) 083537}, [\href{http://arxiv.org/abs/1902.03951}{{\tt
  1902.03951}}].

\bibitem{Kodama:1984ziu}
H.~Kodama and M.~Sasaki, \emph{{Cosmological Perturbation Theory}},
  \href{http://dx.doi.org/10.1143/PTPS.78.1}{\emph{Prog. Theor. Phys. Suppl.}
  {\bf 78} (1984) 1--166}.

\bibitem{Byrnes:2018txb}
C.~T. Byrnes, P.~S. Cole and S.~P. Patil, \emph{{Steepest growth of the power
  spectrum and primordial black holes}},
  \href{http://dx.doi.org/10.1088/1475-7516/2019/06/028}{\emph{JCAP} {\bf 06}
  (2019) 028}, [\href{http://arxiv.org/abs/1811.11158}{{\tt 1811.11158}}].

\bibitem{Salopek:1990jq}
D.~S. Salopek and J.~R. Bond, \emph{{Nonlinear evolution of long wavelength
  metric fluctuations in inflationary models}},
  \href{http://dx.doi.org/10.1103/PhysRevD.42.3936}{\emph{Phys. Rev.} {\bf D42}
  (1990) 3936--3962}.

\bibitem{Sasaki:1995aw}
M.~Sasaki and E.~D. Stewart, \emph{{A General analytic formula for the spectral
  index of the density perturbations produced during inflation}},
  \href{http://dx.doi.org/10.1143/PTP.95.71}{\emph{Prog. Theor. Phys.} {\bf 95}
  (1996) 71--78}, [\href{http://arxiv.org/abs/astro-ph/9507001}{{\tt
  astro-ph/9507001}}].

\bibitem{Wands:2000dp}
D.~Wands, K.~A. Malik, D.~H. Lyth and A.~R. Liddle, \emph{{A New approach to
  the evolution of cosmological perturbations on large scales}},
  \href{http://dx.doi.org/10.1103/PhysRevD.62.043527}{\emph{Phys. Rev. D} {\bf
  62} (2000) 043527}, [\href{http://arxiv.org/abs/astro-ph/0003278}{{\tt
  astro-ph/0003278}}].

\bibitem{Pattison:2019hef}
C.~Pattison, V.~Vennin, H.~Assadullahi and D.~Wands, \emph{{Stochastic
  inflation beyond slow roll}},
  \href{http://dx.doi.org/10.1088/1475-7516/2019/07/031}{\emph{JCAP} {\bf 07}
  (2019) 031}, [\href{http://arxiv.org/abs/1905.06300}{{\tt 1905.06300}}].

\bibitem{Artigas:2021zdk}
D.~Artigas, J.~Grain and V.~Vennin, \emph{{Hamiltonian formalism for
  cosmological perturbations: the~separate-universe approach}},
  \href{http://dx.doi.org/10.1088/1475-7516/2022/02/001}{\emph{JCAP} {\bf 02}
  (2022) 001}, [\href{http://arxiv.org/abs/2110.11720}{{\tt 2110.11720}}].

\bibitem{Grain:2017dqa}
J.~Grain and V.~Vennin, \emph{{Stochastic inflation in phase space: Is slow
  roll a stochastic attractor?}},
  \href{http://dx.doi.org/10.1088/1475-7516/2017/05/045}{\emph{JCAP} {\bf 05}
  (2017) 045}, [\href{http://arxiv.org/abs/1703.00447}{{\tt 1703.00447}}].

\bibitem{Enqvist:2008kt}
K.~Enqvist, S.~Nurmi, D.~Podolsky and G.~Rigopoulos, \emph{{On the divergences
  of inflationary superhorizon perturbations}},
  \href{http://dx.doi.org/10.1088/1475-7516/2008/04/025}{\emph{JCAP} {\bf 04}
  (2008) 025}, [\href{http://arxiv.org/abs/0802.0395}{{\tt 0802.0395}}].

\bibitem{Starobinsky:1986fxa}
A.~A. Starobinsky, \emph{{Multicomponent de Sitter (Inflationary) Stages and
  the Generation of Perturbations}}, {\emph{JETP Lett.} {\bf 42} (1985)
  152--155}.

\bibitem{Lyth:2004gb}
D.~H. Lyth, K.~A. Malik and M.~Sasaki, \emph{{A General proof of the
  conservation of the curvature perturbation}},
  \href{http://dx.doi.org/10.1088/1475-7516/2005/05/004}{\emph{JCAP} {\bf 05}
  (2005) 004}, [\href{http://arxiv.org/abs/astro-ph/0411220}{{\tt
  astro-ph/0411220}}].

\bibitem{Tada:2016pmk}
Y.~Tada and V.~Vennin, \emph{{Squeezed bispectrum in the $\delta N$ formalism:
  local observer effect in field space}},
  \href{http://dx.doi.org/10.1088/1475-7516/2017/02/021}{\emph{JCAP} {\bf 02}
  (2017) 021}, [\href{http://arxiv.org/abs/1609.08876}{{\tt 1609.08876}}].

\bibitem{Ando:2020fjm}
K.~Ando and V.~Vennin, \emph{{Power spectrum in stochastic inflation}},
  \href{http://dx.doi.org/10.1088/1475-7516/2021/04/057}{\emph{JCAP} {\bf 04}
  (2021) 057}, [\href{http://arxiv.org/abs/2012.02031}{{\tt 2012.02031}}].

\bibitem{risken1989fpe}
H.~Risken and H.~Haken, \emph{{The Fokker-Planck Equation: Methods of Solution
  and Applications Second Edition}}.
\newblock Springer, 1989.

\bibitem{2002PASP..114.1051S}
J.~L. {Starck}, E.~{Pantin} and F.~{Murtagh}, \emph{{Deconvolution in
  Astronomy: A Review}}, \href{http://dx.doi.org/10.1086/342606}{\emph{The
  Publications of the Astronomical Society of the Pacific} {\bf 114} (Oct.,
  2002) 1051--1069}.

\bibitem{Shibata:1999zs}
M.~Shibata and M.~Sasaki, \emph{{Black hole formation in the Friedmann
  universe: Formulation and computation in numerical relativity}},
  \href{http://dx.doi.org/10.1103/PhysRevD.60.084002}{\emph{Phys. Rev. D} {\bf
  60} (1999) 084002}, [\href{http://arxiv.org/abs/gr-qc/9905064}{{\tt
  gr-qc/9905064}}].

\bibitem{Harada:2015yda}
T.~Harada, C.-M. Yoo, T.~Nakama and Y.~Koga, \emph{{Cosmological
  long-wavelength solutions and primordial black hole formation}},
  \href{http://dx.doi.org/10.1103/PhysRevD.91.084057}{\emph{Phys. Rev. D} {\bf
  91} (2015) 084057}, [\href{http://arxiv.org/abs/1503.03934}{{\tt
  1503.03934}}].

\bibitem{Choptuik:1992jv}
M.~W. Choptuik, \emph{{Universality and scaling in gravitational collapse of a
  massless scalar field}},
  \href{http://dx.doi.org/10.1103/PhysRevLett.70.9}{\emph{Phys. Rev. Lett.}
  {\bf 70} (1993) 9--12}.

\bibitem{Evans:1994pj}
C.~R. Evans and J.~S. Coleman, \emph{{Observation of critical phenomena and
  selfsimilarity in the gravitational collapse of radiation fluid}},
  \href{http://dx.doi.org/10.1103/PhysRevLett.72.1782}{\emph{Phys. Rev. Lett.}
  {\bf 72} (1994) 1782--1785}, [\href{http://arxiv.org/abs/gr-qc/9402041}{{\tt
  gr-qc/9402041}}].

\bibitem{Niemeyer:1997mt}
J.~C. Niemeyer and K.~Jedamzik, \emph{{Near-critical gravitational collapse and
  the initial mass function of primordial black holes}},
  \href{http://dx.doi.org/10.1103/PhysRevLett.80.5481}{\emph{Phys. Rev. Lett.}
  {\bf 80} (1998) 5481--5484},
  [\href{http://arxiv.org/abs/astro-ph/9709072}{{\tt astro-ph/9709072}}].

\bibitem{Inomata:2020lmk}
K.~Inomata, M.~Kawasaki, K.~Mukaida, T.~Terada and T.~T. Yanagida,
  \emph{{Gravitational Wave Production right after a Primordial Black Hole
  Evaporation}},
  \href{http://dx.doi.org/10.1103/PhysRevD.101.123533}{\emph{Phys. Rev. D} {\bf
  101} (2020) 123533}, [\href{http://arxiv.org/abs/2003.10455}{{\tt
  2003.10455}}].

\bibitem{Papanikolaou:2020qtd}
T.~Papanikolaou, V.~Vennin and D.~Langlois, \emph{{Gravitational waves from a
  universe filled with primordial black holes}},
  \href{http://dx.doi.org/10.1088/1475-7516/2021/03/053}{\emph{JCAP} {\bf 03}
  (2021) 053}, [\href{http://arxiv.org/abs/2010.11573}{{\tt 2010.11573}}].

\bibitem{Domenech:2020ssp}
G.~Dom\`enech, C.~Lin and M.~Sasaki, \emph{{Gravitational wave constraints on
  the primordial black hole dominated early universe}},
  \href{http://dx.doi.org/10.1088/1475-7516/2021/11/E01}{\emph{JCAP} {\bf 04}
  (2021) 062}, [\href{http://arxiv.org/abs/2012.08151}{{\tt 2012.08151}}].

\bibitem{2011ConPh..52..497T}
I.~{Thompson}, \emph{{NIST Handbook of Mathematical Functions, edited by Frank
  W.J. Olver, Daniel W. Lozier, Ronald F. Boisvert, Charles W. Clark}},
  \href{http://dx.doi.org/10.1080/00107514.2011.582161}{\emph{Contemporary
  Physics} {\bf 52} (Sept., 2011) 497--498}.

\end{thebibliography}\endgroup
\end{document}